\documentclass[11pt,a4paper]{article}
\pdfoutput=1 
\usepackage{jinstpub}
\usepackage{placeins}
\usepackage{graphicx}
\usepackage{appendix}
\usepackage{standalone}
\usepackage{subcaption}

\graphicspath{{figures/}}
\DeclareGraphicsExtensions{.pdf,.jpeg,.png,.jpg}

\title{Detector rates for the Small Angle Neutron Scattering instruments at the European Spallation Source}

\author[a,1]{K.~Kanaki%
\note{Corresponding author.}}
\author[a,b,c]{M.~Klausz}
\author[a]{T.~Kittelmann}
\author[d]{G.~Albani}
\author[e]{E. Perelli Cippo}
\author[a,f]{A.~Jackson}
\author[g]{S.~Jaksch}
\author[h]{T.~Nielsen}
\author[b,c]{P.~Zagyvai}
\author[a,i]{R.~Hall-Wilton}

\affiliation[a]{European Spallation Source ESS ERIC, SE-22100 Lund, Sweden}
\affiliation[b]{Hungarian Academy of Sciences, Centre for Energy Research, 1525 Budapest 114., Hungary}
\affiliation[c]{Budapest University of Technology and Economics, Institute of Nuclear Techniques, 1111 Budapest, M\H uegyetem rakpart 9., Hungary}
\affiliation[d]{Dipartimento di Fisica ``G. Occhialini'', Universit\`a degli Studi di Milano-Bicocca, Milan 20125, Italy}
\affiliation[e]{Istituto di Fisica del Plasma ``P. Caldirola'', Associazione EURATOM-ENEA/CNR, Milan 20125, Italy}
\affiliation[f]{Physical Chemistry, Lund University, SE-221 00, Lund, Sweden}
\affiliation[g]{Forschungzentrum J\"ulich GmbH, J\"ulich Centre for Neutron Science JCNS
  at Heinz Maier-Leibnitz Zentrum, D-85747 Garching, Germany}
\affiliation[h]{Data Management and Software Centre (DMSC), European Spallation Source ERIC, 2200 Copenhagen N, Denmark}
\affiliation[i]{Mid-Sweden University, SE-85170 Sundsvall, Sweden}

\emailAdd{Kalliopi.Kanaki@esss.se}

\abstract{Building the European Spallation Source (ESS), the most
  powerful neutron source in the world, requires significant
  technological advances at most fronts of instrument component
  design. Detectors are not an exception. The existing implementations
  at current neutron scattering facilities are at their performance
  limits and sometimes barely cover the scientific needs. At full
  operation the ESS will yield unprecedented neutron brilliance. This  means that one of the most challenging aspects for the new detector
  designs is the increased rate capability and in particular
  the peak instantaneous rate capability, i.e.\,the number of neutrons hitting the
  detector per channel, pixel or cm$^2$ at the peak of the neutron pulse. This
  paper focuses on estimating the incident and detection rates that are anticipated
  for the Small Angle Neutron Scattering (SANS) instruments planned
  for ESS. Various approaches are applied and the results thereof are
  presented.}

\keywords{BAND-GEM, Boron Coated Straws, Boron10, detector rate,
  Geant4, He3, McStas, neutron detector, neutron scattering, SoNDe, transmission monitor}

\begin{document}
\maketitle 


\section{Introduction}

The European Spallation Source~\cite{esscdr,esstdr} aspires to become the most powerful
neutron source in the world, presenting new challenges for the
detector technologies to be used. Fifteen neutron instruments that have
been approved as part of the ESS instrument suite are already challenging the
traditional detection technologies of the neutron scattering field because of
the unprecedented instantaneous counting rates, as well as the
requirements for high spatial resolution~\cite{kirstein2014}.

LoKI~\cite{lokijackson2015,lokiproposal} and SKADI~\cite{skadi2014,skadi2016} are the two Small Angle Neutron Scattering (SANS) instruments endorsed for
construction at ESS. LoKI is designed to be a broad-band high intensity
instrument with wide polar angle coverage. SKADI offers a
broad range of neutron energies as well, in addition to neutron polarisation but with a
smaller angular coverage, with its detectors placed at a larger distance from
the sample compared to LoKI.

In order to design detectors with the appropriate rate capability and prevent performance
compromises on this front, a detailed analysis of the instrument
requirements and how they translate to detector requirements is
vital. It is also important that such an evaluation be performed with
different and complementary approaches, so as to ensure the
reliability of the result within acceptable limits.

The methods used in this paper aim at acquiring a good
understanding of the rates LoKI and SKADI are going to achieve by
looking at how the scattered neutrons on the sample are distributed
in space and time. A realistic worst-case scenario is reproduced, based
on which the detector requirements for rate capability are extracted
and the choice of detector technology is discussed.

In the following sections the tools and methods used are elaborated upon. The main focus of this
evaluation concerns the neutron scattering taking place within the forward solid
angle after the sample, as this happens to be the primary area of interest for the SANS
technique.
This translates to a typical 1~m~$\times$~1~m detector area
considered for the rate estimates of the scattering characterisation
system (SCS). Various detector technologies are explored and their traits are
discussed with respect to the instrument rate requirements, as well as
the data acquisition (DAQ) design. Last but not least, a
smaller area detector acting as a transmission monitor is placed in the
path of the direct beam and the same analysis is repeated.

\section{The rate challenge for the ESS SANS detectors}

SANS is one of the most widely used
experimental techniques for the investigation of soft matter
properties~\cite{sansbook}. It uses neutrons that are elastically scattered on the
sample at small scattering angles to infer the characteristic length scales of the structures present in
it. The scale of the structures of interest that SANS can access is
1--100~nm. Its use of contrast variation from deuterated samples makes
it a very powerful tool especially for studies of
biological and pharmaceutical interest. 

The popularity of the SANS technique makes it a ubiquitous tool at
every neutron scattering facility, usually with multiple instruments,
targeting both at high availability for the scientific community, as
well as satisfying complementary scientific requirements. For the same
reason, two SANS instruments have been endorsed at ESS and are
currently undergoing different stages of design.


\subsection{The LoKI instrument}
LoKI (the Low-K Instrument) is a wide simultaneous Q range SANS instrument designed
primarily with the needs of the soft matter, biophysics and materials science
communities in mind. The trend in all of these fields is towards complexity and
heterogeneity. Multi-component systems need to be studied as a
function of multiple environmental conditions with different structures
occurring at different length scales. Small gauge volumes are important to study
both intrinsically small samples and for performing scans of heterogeneous samples.
There is a need to study systems under the sort of non-equilibrium conditions,
such as shear fields, found in real world applications. The ESS flux will enable
more routine access of these conditions and also permit fast kinetic studies on a 
wider range of samples than is possible today. These scientific requirements
motivate the requirements that the detector system have a large angular coverage,
have good spatial resolution, and have a high rate capability.

\subsection{The SKADI instrument}

The Small-K Advanced DIffractometer SKADI is
a high-resolution SANS instrument at the ESS. The scientific areas targeted by SKADI
include investigations of smart materials, biological and medical
research, magnetic materials and materials for energy storage as well
as experiments on nano-materials and nano-composites. As an additional
feature SKADI offers a modular sample environment (3~m~$\times$~3~m)
that allows custom sample environments, especially for in-situ and
in-operando investigations. Those will greatly benefit from the higher
flux at the ESS due to lower accessible time scales. In order to make
full use of the higher flux, a high-rate capable detector, such as the
Solid-state Neutron Detector (SoNDe), is necessary.

\begin{figure}
\centering
\includegraphics[width=0.6\textwidth]{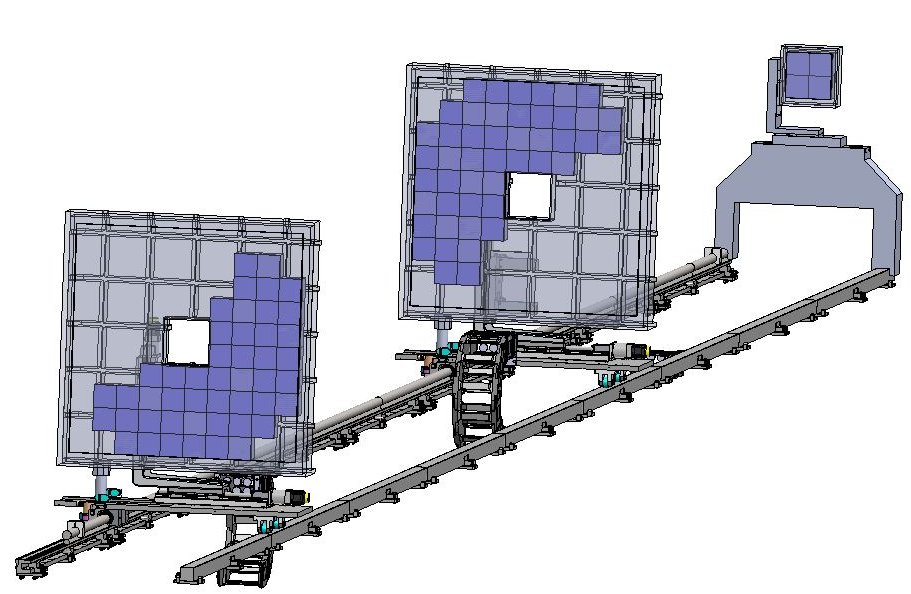}
\caption{Arrangement of three SoNDe detector banks for the SKADI
  instrument. The single modules for the active area are
  indicated. 
}
\label{sonde_drawing}
\end{figure}

The detector setup includes three detector banks at different
positions (see Fig.~\ref{sonde_drawing}). One detector of
20~cm~$\times$~20~cm will be installed at a fixed position 20~m from
the sample. The central detector is a 1~m~$\times$~1~m with a central
20~cm~$\times$~20~cm aperture that will always be at the collimation
distance to allow for ideal mapping of resolution between aperture and
detector. An identical detector is placed at 1/5 of the middle
detector distance from the sample to allow collecting data from higher
scattering angles. This setup allows covering three orders of
magnitude in the Q-space (reciprocal momentum transfer in the neutron interaction~\cite{colin}, p.\,26) in a single shot experiment to enable the observation of fast sample transitions.


\FloatBarrier

\subsection{Instrument configurations}

The chopper settings, collimation length and aperture sizes of a SANS
instrument are experimentally configured depending on the sample and
the type of measurement to be performed. For the rate estimates a few
configurations are selected, aiming at maximising the neutron flux on
the sample, in combination with a sample that strongly scatters within
the forward solid angle. 

The instrument and sample settings used for the rate estimates are listed in
Tab.~\ref{instr_config_table}. The chopper settings are reflected in
the wavelength range, i.e.\,the $\lambda_{min}$ and $\lambda_{max}$
values allowed to reach the sample from the source. The
same sample model is used for all instrument configurations. It is
a diblock copolymer with a 4.84\% scattering probability, an adequate
choice for the study of the rear detector geometry. The sample size of
1~cm~$\times$~1~cm matches that of the sample aperture.
\begin{table}[h]
  \centering
  \caption{\footnotesize LoKI instrument configurations for the
    evaluation of rates. The same sample is used for all
    scenarios. The divergence represents the direct beam size on the
    detector and is defined by the collimation
    settings.} 
  \begin{tabular}{|c|c|c|c|c|c|c|c|}
    \hline
    config & collimation & $\lambda_{min}$ & $\lambda_{max}$ & source
    & sample & flux on & divergence \\
           & length  &           &          & aperture
    & aperture & sample & \\
    & (m) & (\AA) & (\AA) & (cm~$\times$~cm) & (cm~$\times$~cm) &
    (n/cm$^2$/s) & (cm) \\
  \hline
  1 & 3 & 3.0 & 11.5 & 3~$\times$~3 & 1~$\times$~1 & 1.00$\times$10$^9$ &8.4 \\ \hline
  2 & 5 & 3.0 & 11.5 & 2~$\times$~2 & 1~$\times$~1 & 2.02$\times$10$^8$ &4.4\\ \hline
  3 & 8 & 3.0 & 10.0 & 2~$\times$~2 & 1~$\times$~1 & 8.24$\times$10$^7$ &3.0\\ \hline
  \end{tabular}
  \label{instr_config_table}
\end{table}
In SANS it is typical to match the collimation length to the sample-detector
distance, in order to optimise the resolution capability of the
instrument. For the current study though, this distance is fixed at
5~m to get a cross comparison among the detector options presented
in the following sections.

The contributions to the number of neutrons reaching the detector can
be broken down to the coherent and incoherent scattering components
from the sample, the incoherent scattering from the solvent and the
transmitted neutrons, as in Eq.~\ref{SB}.
\begin{equation}
N = coh_{sample} + incoh_{sample} + incoh_{solvent} + transmission.
\label{SB}
\end{equation}
In the current study, only the coherent scattering from the sample is
taken into account, i.e.\,the actual signal the SANS technique is
after. This provides a lower limit to the rate estimates for the SCS. The transmission neutrons are studied
separately to shed light on the requirements of the direct beam
detector. In the latter case the derived rates serve as an upper
limit.



\subsection{Rate definitions and analytical estimates}

The rate definitions used in this study are listed in~\cite{diffrrates} (page 14) as follows:
\begin{itemize}
\item \textbf{Global time-averaged incident/detection rate}:  the number of
  neutrons per second entering/recorded by the entire detector. 
\item \textbf{Local time-averaged incident/detection rate}: the number of neutrons
  per second entering/recorded in a detector pixel, channel or unit.  
\item \textbf{Global peak incident/detection rate}: the highest
  instantaneous neutron incident/detection rate on the whole detector.
\item \textbf{Local peak incident/detection rate}: the highest
  instantaneous neutron incident/detection rate on the brightest detector pixel, channel or unit. 
\end{itemize}
Knowing the scattering fraction F(\%) for a SANS sample model, it is
possible to get an approximate analytical estimate for the time-averaged incident rates
the rear detector is exposed to. The largest portion of the signal-related scattered
neutrons is contained within the forward cone subtended by the rear
detector, which usually has a polar angle coverage of up to a few
degrees, depending on its distance from the sample. As SANS deals with different degrees of freedom compared to
other neutron techniques, the neutrons scattered at small angles
represent long range correlations and carry information about large
distances in real space. The diblock copolymer model mentioned before
has a scattering fraction of about 4.8\%. For instrument configuration
1 this translates to an upper limit of a global incident time-averaged signal rate of
4.8\%~$\times$~10$^9$n/s/cm$^2$~flux~$\times$~1~cm$^2$ sample size =
4.8~$\times$~10$^7$~n/s = 48 MHz. The incoherent
contribution from the sample and the solvent have to be estimated in
addition. If 10\% of the incident beam is incoherently scattered in
4$\pi$, then the incident fraction within the solid angle subtended by
the rear detector is 10\%~$\times$~10$^9$n/s/cm$^2$~flux~$\times$~1~cm$^2$
sample size $\times$ 0.04~sr = 4~MHz. A global time-averaged rate of 52~MHz
is to be expected as an upper incident rate limit for the high flux instrument
configuration on the rear detector.

\subsection{The Monte Carlo approach}

Analytical rate estimates are based on several approximations and lack the more detailed picture that
a Monte Carlo (MC) approach can provide. One fundamental MC tool that is
broadly used for instrument design is McStas~\cite{mcstas1,
  mcstas2}. It can model all major components of an instrument by sampling
an external neutron source distribution and then propagating neutrons through guides,
choppers, monochromators, collimators etc.\,until the sample and beyond. A collection of sample
components is also available allowing the neutrons to scatter based on
underlying physics models and then be histogrammed by monitor
components that act like ``detectors''. The latter process is of
particular interest for the detector design, as it allows the
visualisation of the neutrons entering the detector geometry accessing at the same time
their full list of properties, e.g.\,wavelength ($\lambda$),
time-of-flight (TOF), position, scattering angle ($\theta$) and optionally user-defined flags. These properties
are necessary to differentially study the picture of the impinging
neutrons on the detector as a function of time (within a neutron pulse),
wavelength and polar angle. 

Both ESS SANS instruments are modeled in McStas as part of the design
progression. In this paper, a baseline LoKI model is used assuming that no
major modifications are anticipated or that future modifications will
not impact the predicted rates in a way that completely
resets the detector design effort. The deduced rate numbers correspond
to the butterfly moderator source (``ESS\_butterfly'')~\cite{essmoderator} with a pulse
length of 2.86~ms and reflect the maximum accelerator power of 5~MW.

\subsection{Rate derivation method}
As the neutron scattering community is transitioning from the
traditional $^3$He detectors to solid converter detectors and more
exotic geometries, it becomes important to examine incident and
detection rates for all the detector technologies presently evaluated. These
scenarios include $^3$He tubes, Boron Coated Straws (BCS) by Proportional
Technologies, Inc., the SoNDe detector and the BAND-GEM detector
(refs.\,in respective sections). This
practically translates to a variety of channel and pixel geometries,
from the familiar two-channel tube read out in charge division
mode, to two-dimensional (2D) pixels of rectangular or trapezoidal shape
with individual channel readout.

The output of the McStas (2.4.1) 
simulation that is recorded immediately after the sample is used as input for
a subsequent
Geant4~\cite{geant4a,geant4b,geant4c_inpresscorrectedproof} detector
simulation. The latter is performed with the ESS Detector Group
Simulation Framework~\cite{dgcode}. The file format facilitating this
communication is the Monte Carlo Particle Lists (MCPL)
format~\cite{mcplpaper,mcplgithub}. With the combination of the two
software packages it is possible on one hand to take into account the
correlations of neutron properties, like TOF, energy, spatial and momentum vectors, and on
the other hand to reliably evaluate the detector performance. The
Geant4 validation of the latter is addressed elsewhere~\cite{in6cncs} and is not
discussed in this study.

At a pulsed source the peak rate could be more than an order of
magnitude higher than the time-averaged one. It is thus important for
the qualification of a detector technology for a specific application
and impacts the readout electronics design. The peak rates also
dictate the system design for data collection, aggregation and
transfer but as the data are propagated further down the
process chain, it is the time-averaged rates that become more relevant
for data transfer and storage choices. 

Another distinction in the current work is between incident and
detection rate. The first would be the upper limit the detector would
count, if it had a detection efficiency of 100\%. As the detectors
have efficiencies that depend on the neutron energy, the instance when the number of incident neutrons is highest is not necessarily
the instance of the highest number of detected neutrons. A detector
might be illuminated with a high flux but if it is not efficient at
the respective neutron wavelengths, there is no concern for
saturation. The peak incident rate and peak detection rate do not have
to occur simultaneously.

The incident neutrons arrive at the detector with various TOF and energy
values. As Geant4 continues to count time for the particles it
propagates, the TOF distribution will vary as a function of detector
depth. This effect needs to be taken into account for the estimate of the
instantaneous peak detection rates for those detectors that record
depth information~\cite{in6cncs,piscitelli2017,mgcncs,lacy2013}.

\begin{figure}[!h]  
  \centering
  \begin{subfigure}{0.5\textwidth}
    \centering
    \includegraphics[width=\textwidth]{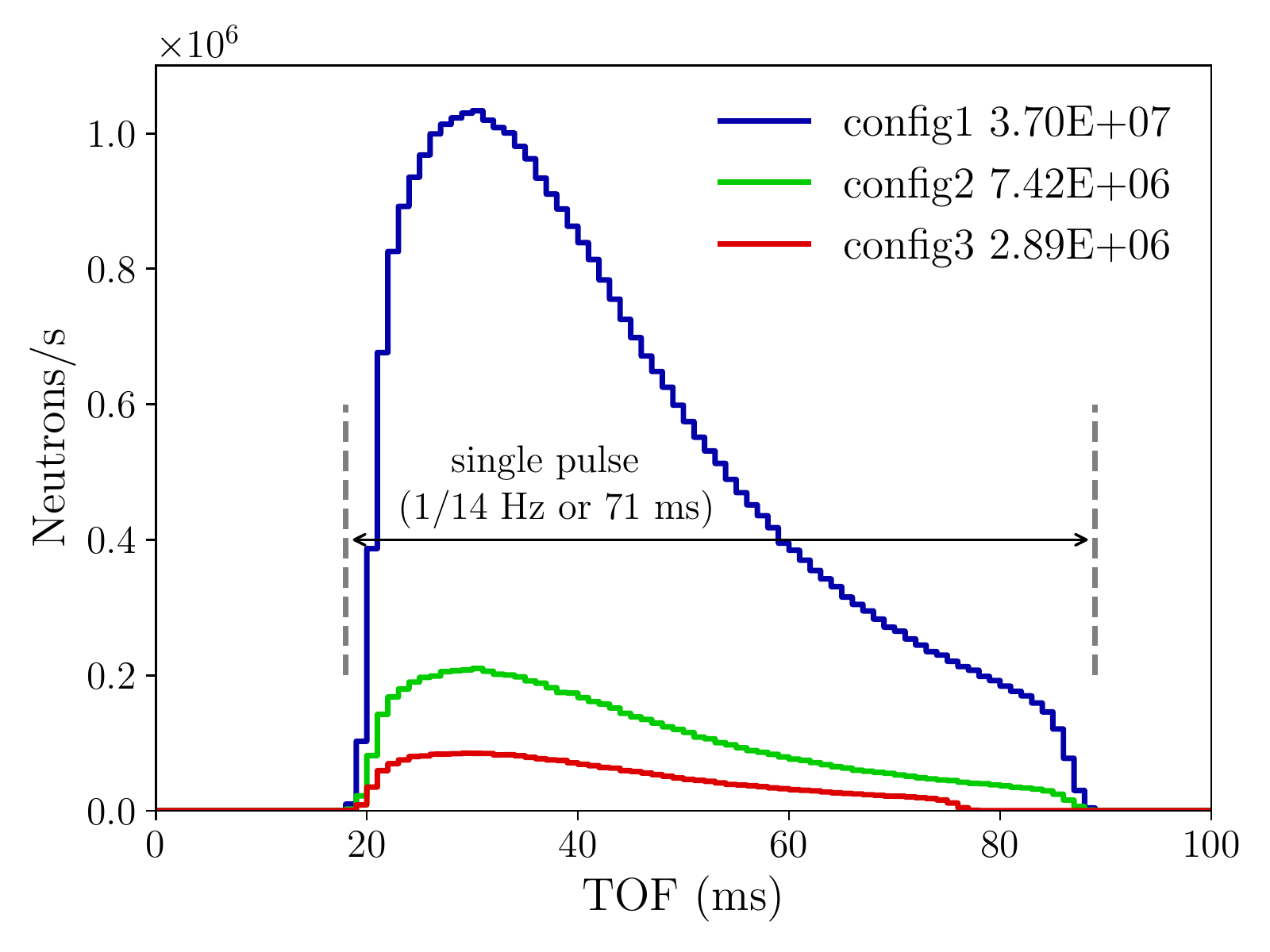}
    \label{tof}    
  \end{subfigure}%
  \begin{subfigure}{0.5\textwidth}
    \centering
    \includegraphics[width=\textwidth]{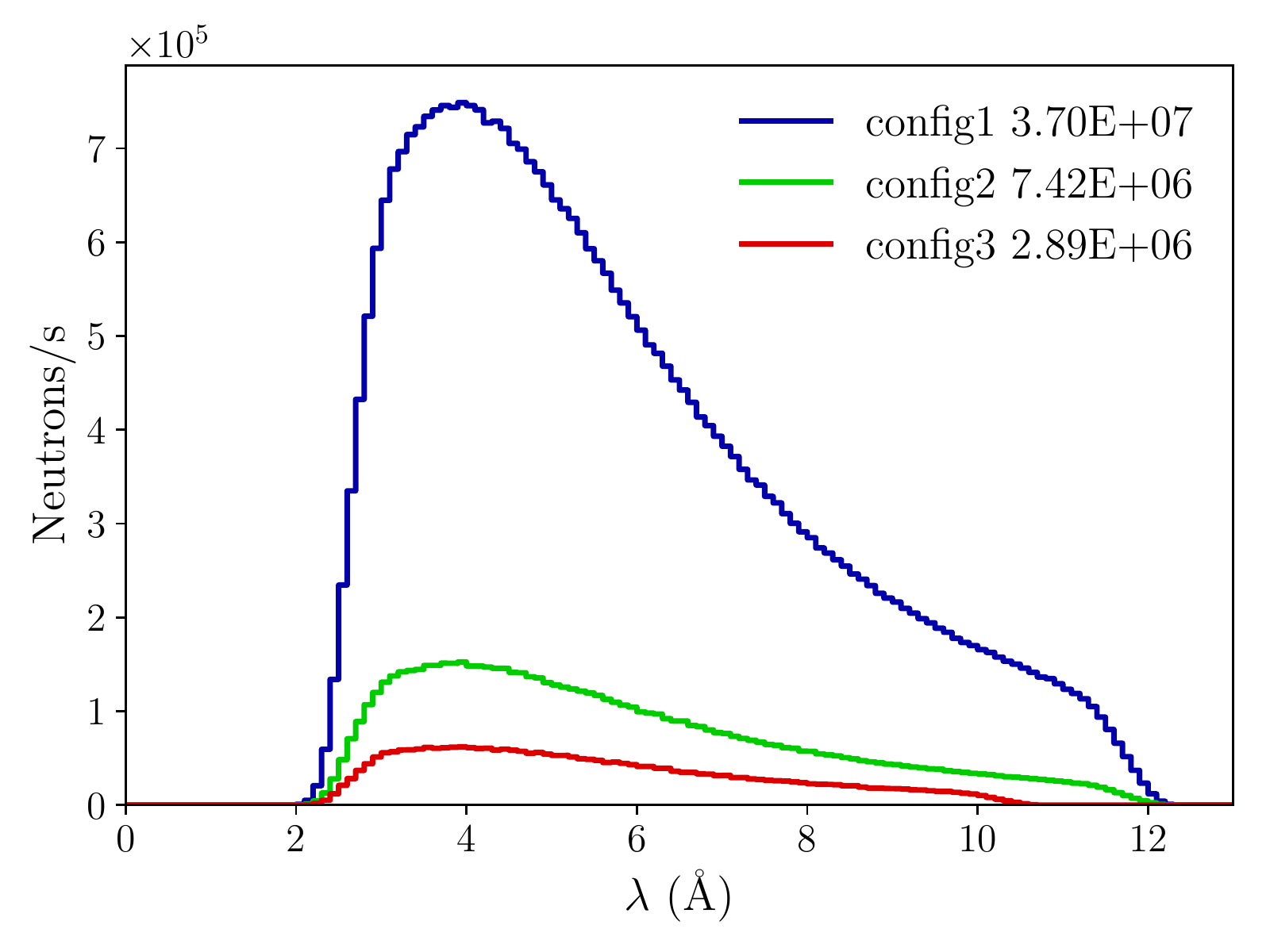}
    \label{lambda}
  \end{subfigure}
  \caption{\footnotesize A typical neutron TOF and $\lambda$ distribution 5~m after the
sample for the three instrument configurations. The global incident rates per
configuration appear in the legend in Hz for a standard 1~m~$\times$~1~m detector.}
  \label{tof_lambda}
\end{figure}

For the derivation of the incident rates, the incident
neutron TOF distribution at 5~m after the sample is used, as depicted
in Fig.~\ref{tof_lambda}, in order to select the TOF region which contains
the highest number of neutrons. Similarly, the maximum of the TOF
distribution of detected neutrons is used for the estimate of the
highest detection rate. 

The peak number of neutrons can be represented in various ways, e.g.\,per tube or
pixel. When filling the histograms representing the peak rates, the weight of every
bin is normalised with the accelerator frequency (14~Hz). 
\FloatBarrier

\section{Tackling the ESS flux with tube geometries}

One of the options the neutron scattering community has primarily relied
on is $^3$He tube detectors arranged in a single layer, typically with
an 8~mm outer diameter and charge
division readout. To satisfy the high rate requirements the tubes can
be arranged in successive layers of possibly lower
detection efficiency, $^3$He or other technology, in order to distribute
the detection events in a larger volume and reduce event
pile-up. The same can be achieved by reducing the tube size but the
increase of dead area compromises the data quality.

\subsection{Rates for the $^3$He tubes} \label{sec:he3}

\begin{figure}[!h]  
  \centering
  \begin{subfigure}{7cm}
    \includegraphics[width=11cm]{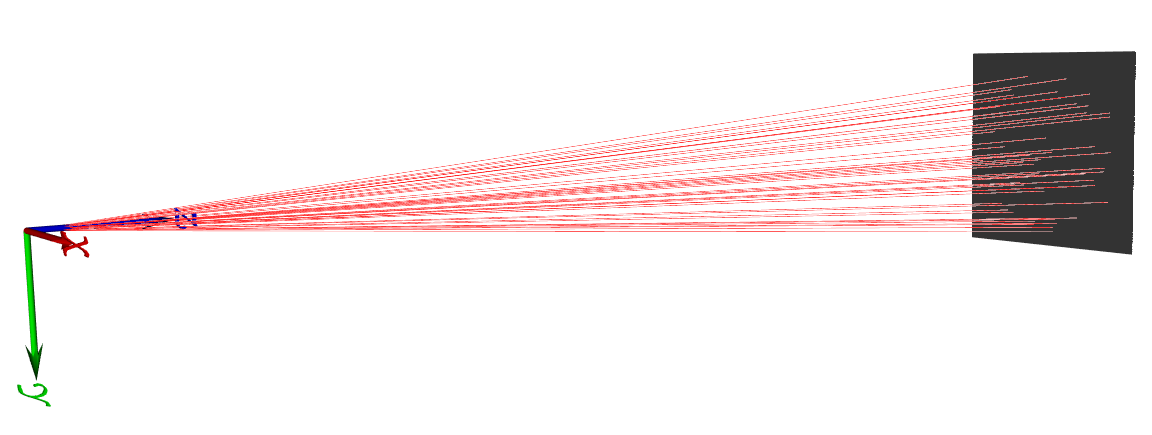}
    \caption{\footnotesize The 1~m~$\times$~1~m detector area covered with 120
      $^3$He tubes. Primary neutrons stemming from the sample position appear in red.}
    \label{he3geo_neutrons}    
  \end{subfigure}%
  \begin{subfigure}{7cm}
    \raggedleft
    \includegraphics[width=2cm]{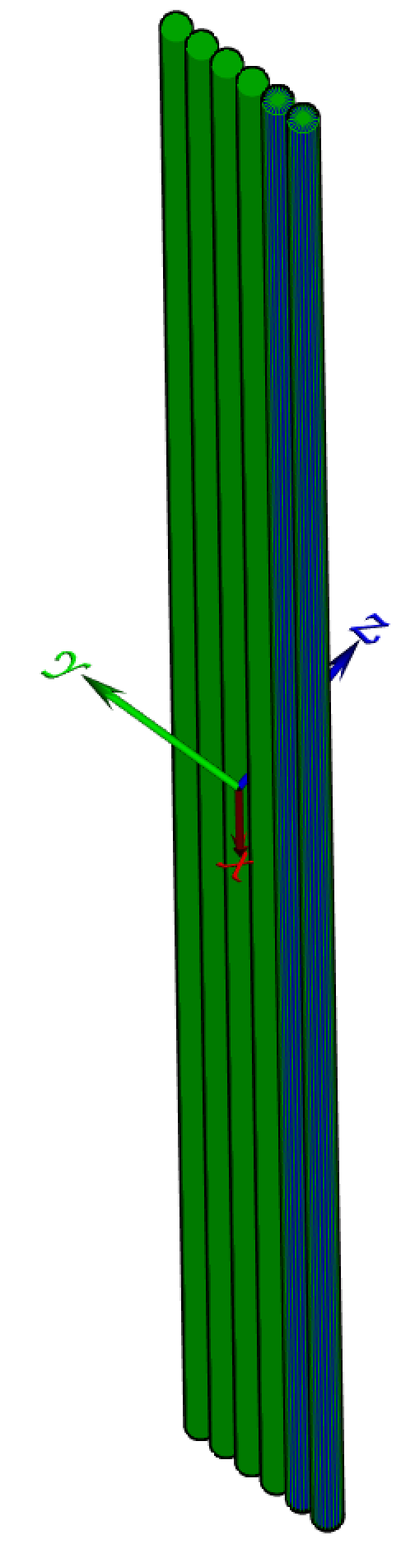}
    \caption{\footnotesize Enlarged view of $^3$He tubes. The stainless steel vessel appears in green and the counting gas in blue.}
    \label{he3geo_zoom}
  \end{subfigure}
  \caption{\footnotesize The neutron generator is placed at the
    beginning of the coordinate system, which coincides with the
    sample position. Simulated neutrons from McStas are emitted
    towards the 1~m$^2$ geometry at 5~m away along the z-axis. Only
    neutrons hitting the detector are displayed.}
  \label{he3geo}
\end{figure}

The Geant4 geometry model implemented for the $^3$He tube (see Fig.~\ref{he3geo}) constitutes a typical detector
arrangement for many SANS instruments around the
world~\cite{sans1,sans2d,eqsans,d33}. The outer tube diameter is 8~mm with a 0.4~mm wall thickness, a length of
1~m and a total gas pressure of 10~bar with a $^3$He/CF$_4$ mixture
(80/20 by volume)~\cite{reuterstokes,toshiba,illmultitube}. The material of the vessel is
stainless steel with the crystalline structure of $\gamma$--iron and
is implemented with the help of the NXSG4 library~\cite{nxsg4}. The
physics list used is QGSP\_BIC\_HP. In addition, \texttt{/process/eLoss/stepFunction 0.1 0.001~$\mu$m} and
\texttt{/process/eLoss/minKinEnergy 10 eV} are defined. 120 tubes are placed at a 5~m distance from the sample position, in
order to achieve a polar angle coverage of a few degrees.

\begin{figure}[!t]  
  \centering
  \begin{subfigure}{0.5\textwidth}
    \centering
    \includegraphics[width=\textwidth]{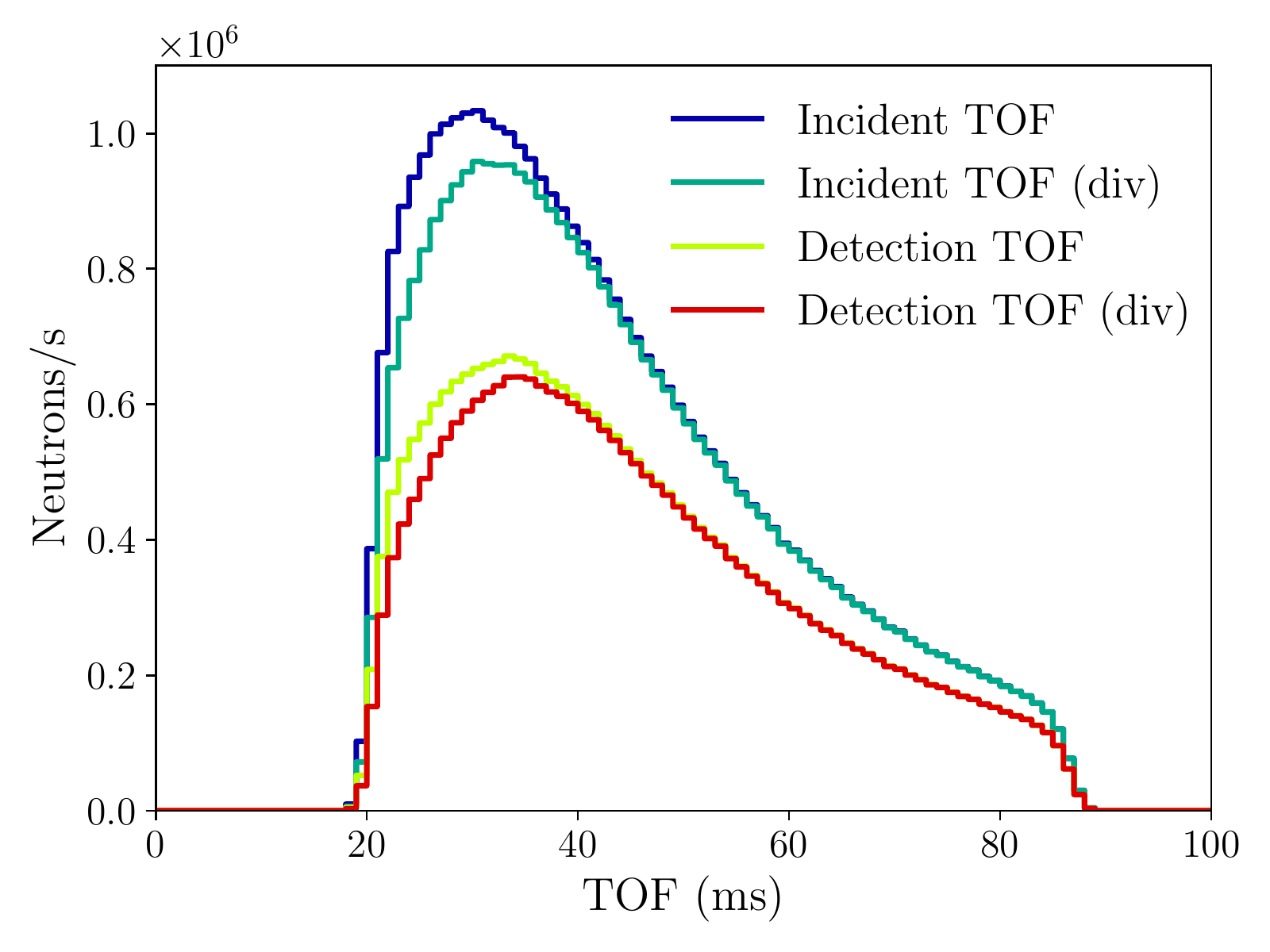}
    \label{he3_tof}    
  \end{subfigure}%
  \begin{subfigure}{0.5\textwidth}
    \centering
    \includegraphics[width=\textwidth]{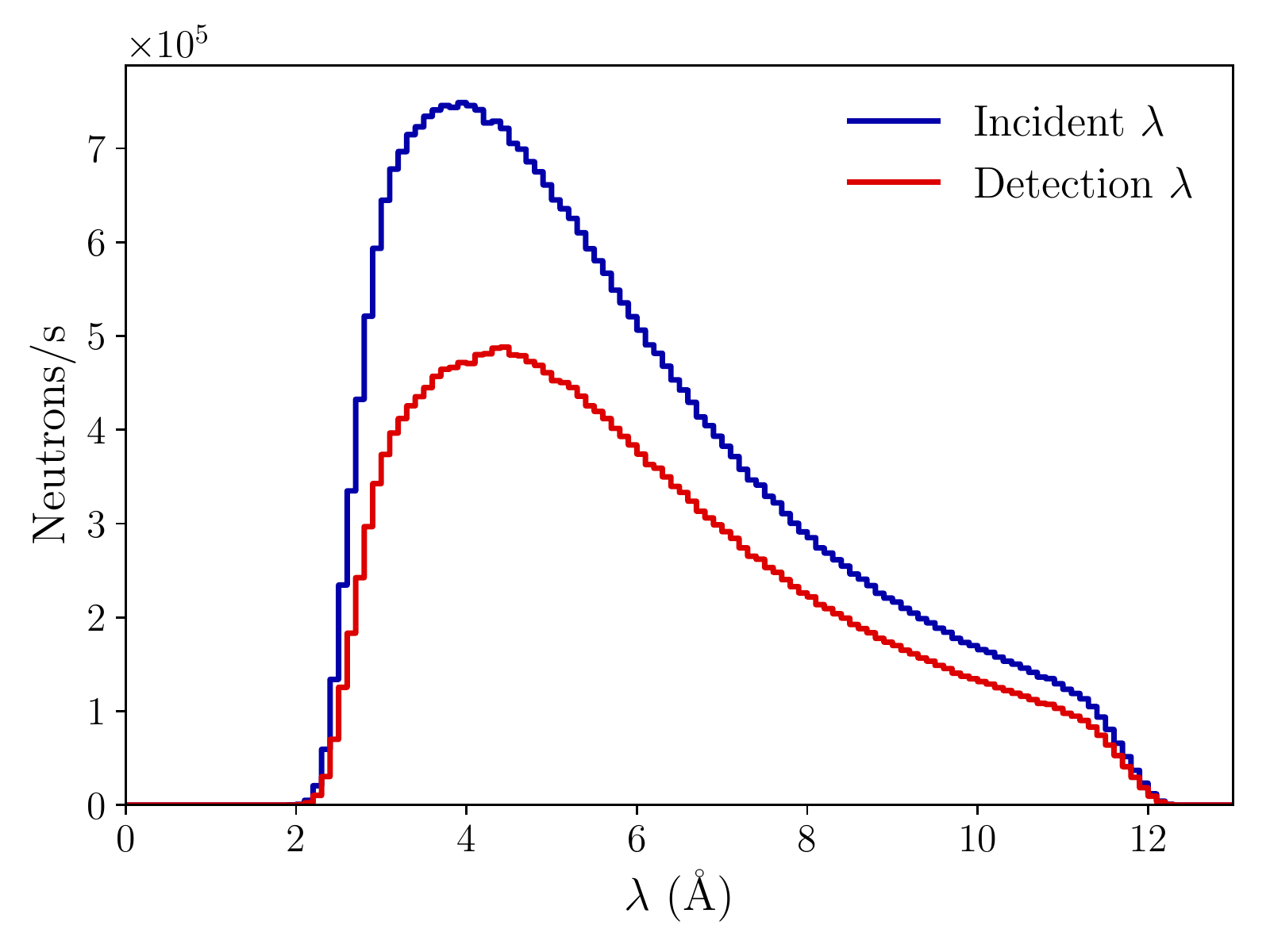}
    \label{he3_lambda}
  \end{subfigure}
  \caption{\footnotesize Incident and detected neutron TOF
    distributions for instrument configuration 1, including and
    excluding divergent neutrons. The TOF maxima can appear at different
    values for the two cases (left). Incident and detected neutron $\lambda$ distributions for the same instrument configuration (right).}
  \label{he3_tof_lambda}
\end{figure}

A neutron is counted as incident, when it enters the wall material of
a tube. It is counted as detected, when its conversion products have
deposited more than 120~keV in the counting gas. The threshold is based on the shape of
the experimental pulse height spectrum and has a minor impact on the result of this study. The TOF and
wavelength distributions of incident and detected neutrons for
instrument configuration 1 are shown in Fig.~\ref{he3_tof_lambda}. For
every evaluation that follows the TOF maximum is picked from the
respective distribution.  

\begin{figure}[!h]  
  \centering
  \begin{subfigure}{0.5\textwidth}
    \centering
    \includegraphics[width=\textwidth]{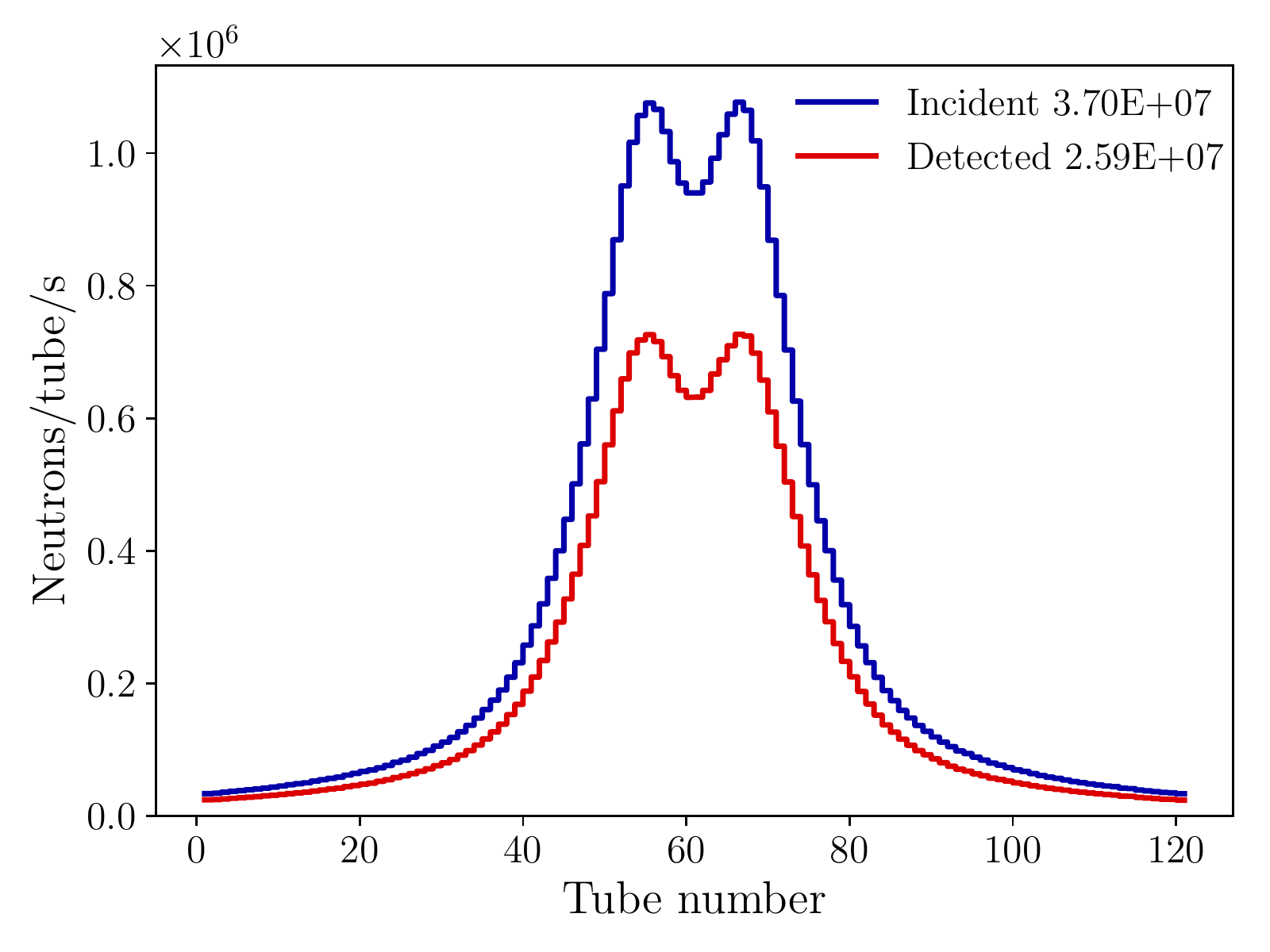}
    \caption{\footnotesize Time-averaged rates per $^3$He tube.}
    \label{he3_average}    
  \end{subfigure}%
  \begin{subfigure}{0.5\textwidth}
    \centering
    \includegraphics[width=\textwidth]{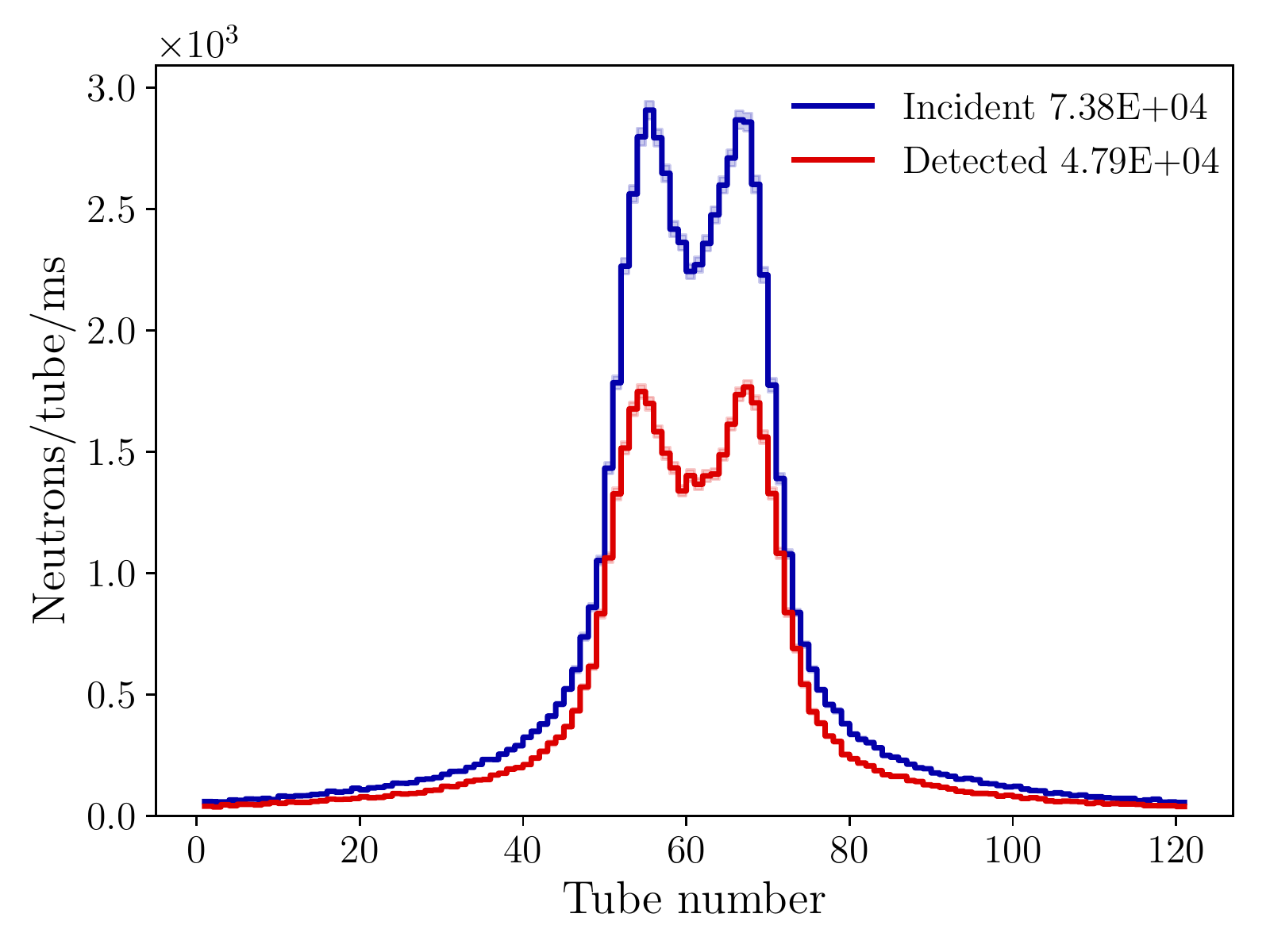}
    \caption{\footnotesize Peak rates per $^3$He tube.}
    \label{he3_peak_rate}
  \end{subfigure}
  \caption{\footnotesize Global (in legend) and local, incident and detection
    rates for a $^3$He detector and instrument configuration 1.}
  \label{he3_rates}
\end{figure}

Fig.~\ref{he3_average} depicts the time-averaged global and local
rates for both incident and detected neutrons. The number cited in the
legend corresponds to the global time-averaged rate, while the maximum of
each histogram corresponds to the local one. In order to extract the peak rates, a 1~ms TOF slice is selected from
the respective distribution and only neutrons within this slice are
plotted (see Fig.~\ref{he3_peak_rate}). The counts are per tube and ms
but are converted to Hz when cited in the summary
Tab.~\ref{he3_rate_table}. The inclusion or exclusion of divergent
neutrons amounts to a 5\% difference only for the first configuration
and the particular sample model, and as such it is not presented
as a separate case.
The same analysis is repeated for all instrument configurations of
Tab.~\ref{instr_config_table} and the results are summarised in
Tab.~\ref{he3_rate_table}.  
\begin{table}[h]
  \centering  
  \caption{\footnotesize Summary table with global and local incident and detection
    rates, both time-averaged and peak ones, for $^3$He tubes.}
  \begin{tabular}{|c|c|c|c|c|}
    \hline
    config & global average   & global average    & local average                  & local average  \\
    & \textbf{incident} rate & \textbf{detection} rate & \textbf{incident} rate/tube & \textbf{detection} rate/tube \\
    \hline
    1 & 37~MHz  & 25.9~MHz & 1~MHz   & 727~kHz \\ \hline
    2 & 7.4~MHz & 5.2~MHz  & 225~kHz & 152~kHz \\ \hline
    3 & 2.9~MHz & 2~MHz    & 91~kHz  & 61~kHz  \\ \hline
     \hline
    config & global peak   & global peak    & local peak                  & local peak  \\
    & \textbf{incident} rate & \textbf{detection} rate & \textbf{incident} rate/tube & \textbf{detection} rate/tube \\
    \hline
    1 & 74~MHz & 48~MHz  & 2.9~MHz & 1.8~MHz \\ \hline
    2 & 15~MHz & 9.6~MHz & 647~kHz & 381~kHz \\ \hline
    3 & 6~MHz  & 3.9~MHz & 279~kHz & 175~kHz \\ \hline
  \end{tabular} 
  \label{he3_rate_table}
\end{table}

The operational limit for a non position sensitive $^3$He tube is
about 100~kHz~\cite{knollhe3,illbb}, dropping at 30-50~kHz for
position sensitive tubes. At higher rates the tube performance
deteriorates leading to event pile-up. The spatial resolution from
charge division is sensitive to the position of the detection event
along the tube and is additionally subject to compromise when the
operation parameters are favoured towards high rate capability. 

The $^3$He technology is clearly inappropriate for the SANS
instruments at ESS. Alternative solutions need to be adopted that can
serve the scientific case without wasting the intense ESS neutron
pulse. A detector option that resembles the $^3$He tube readout but
uses a solid $^{10}$B$_4$C converter is explored next.  
\FloatBarrier

\subsection{Rates for the Boron-Coated Straws}

\begin{figure}[!h]  
  \centering
  \begin{subfigure}{0.40\textwidth} 
    \includegraphics[width=0.9\textwidth]{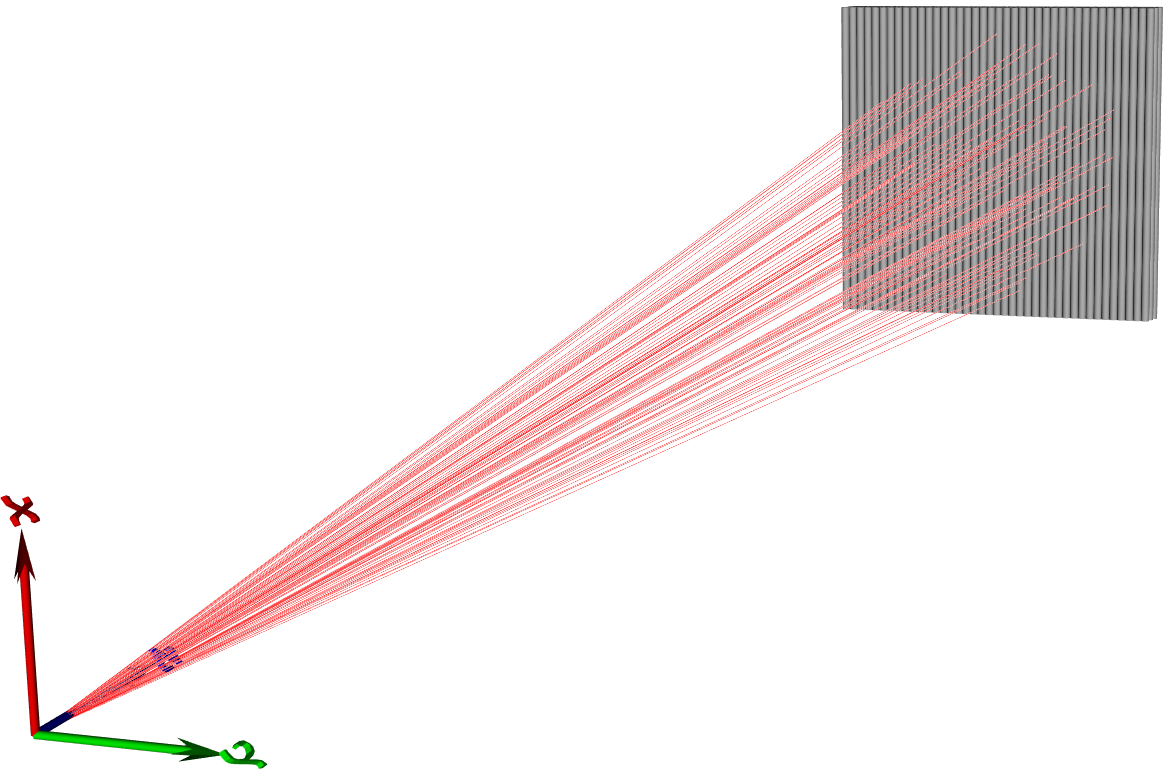}
    \caption{\footnotesize The primary neutrons (in red) hit
      the detectors 5~m away from the sample position.}
    \label{bcsGeo_neutrons}    
  \end{subfigure}
  \begin{subfigure}{0.59\textwidth} 
    \centering
    \includegraphics[width=\textwidth]{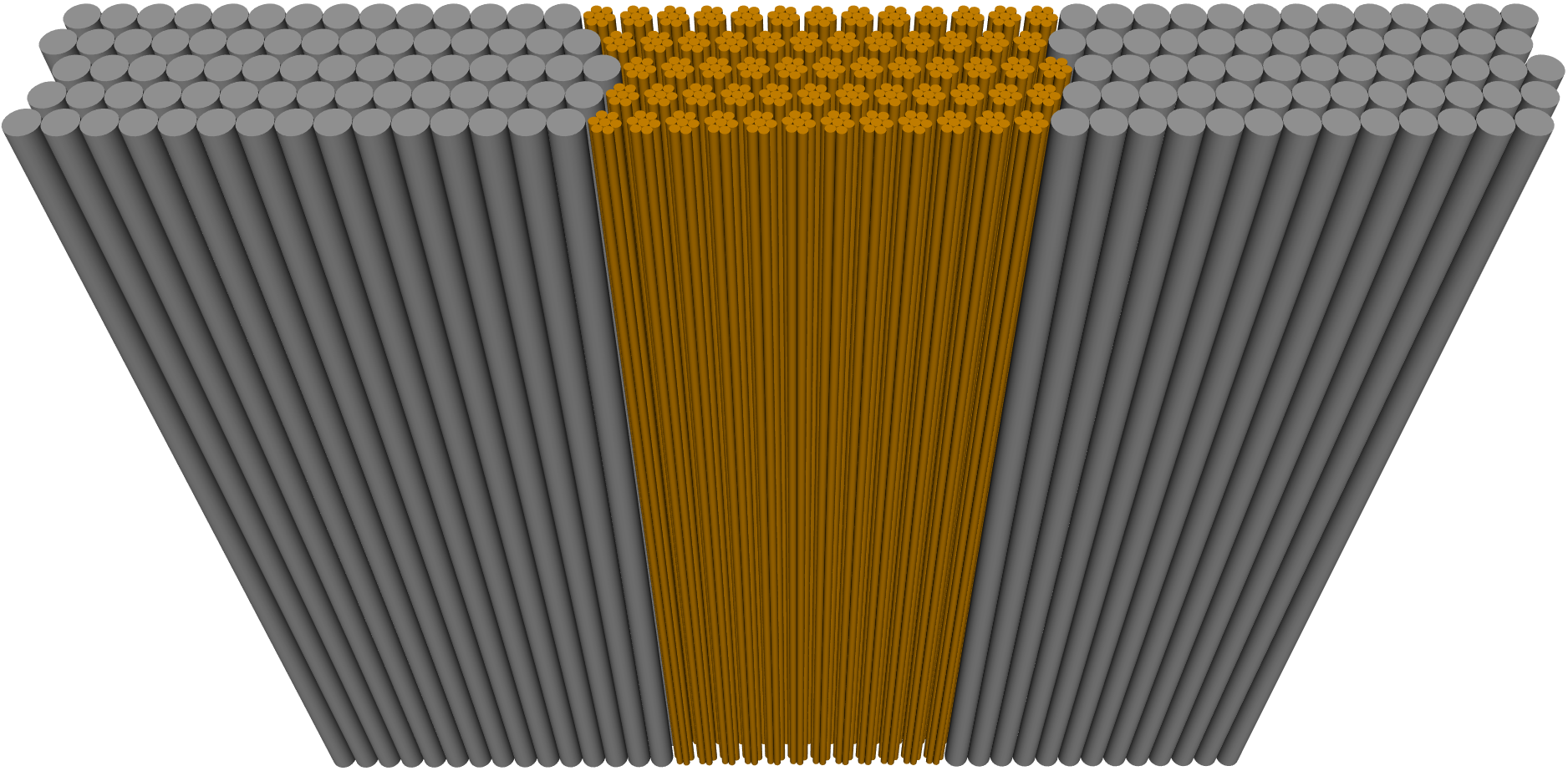}
    \caption{\footnotesize Enlarged view of the 5 overlapping BCS tube
      panels. Each aluminium tube (grey) contains 7 copper straws (orange) arranged hexagonally.}
    \label{bcs_zoom}
  \end{subfigure}
  \caption{\footnotesize The neutron generator is placed at the
    beginning of the coordinate system, which coincides with the
    sample position. Simulated neutrons from McStas are emitted
    towards 5 panels ($\times$~40 tubes/panel) of BCS tubes covering 1~m~$\times$~1~m at 5~m away along the
    z-axis.}
  \label{bcsGeo}
\end{figure}

A Boron-Coated Straw (BCS)~\cite{boronstraw_lacy_2010, lacy2013} detector is a conventional neutron detector
by Proportional Technologies, Inc., which intends to be
a cost effective $^3$He replacement technology.
A BCS detector consists of an aluminium tube, containing seven
copper straws arranged hexagonally (see Fig.~\ref{bcs_zoom}). The tubes are 1~m long with a
diameter of 2.54~cm. The straw inner wall is
coated with a 1~$\mu$m thin B$_4$C converter layer 
enriched in $^{10}$B by 95\%. They are filled with an Ar/CO$_2$
mixture (90/10 by volume) at 0.7~atm. Physics list and \texttt{eLoss}
parameters are identical to the $^3$He tube simulation (see
introduction of section~\ref{sec:he3}). The neutron transport in the
Al and Cu materials is described with the help of the NCrystal package~\cite{ncrystalgithub, icns}.

A bias voltage is applied between the tube and resistive
anode wires, which are tensioned in the center of each straw.
This makes the straws work in proportional mode. The charge is read out at both ends
of the detector using charge division to acquire the position information. The
tubes can be arranged in successive layers in order to achieve the
desired coverage, uniformity and detection efficiency.



In the Geant4 model of the current study 5 layers of BCS detectors, referred to as
panels, consisting of 40 tubes each cover the same 1~m~$\times$~1~m area as the $^3$He detector model. Each tube
is rotated by an angle of 10$^{\circ}$ around its cylindrical axis and the adjacent panels are
positioned behind each other with a relative shift of 1.016~cm along
the horizontal axis (see
Fig.~\ref{bcs_zoom}) for performance optimisation reasons. 

A neutron detection event is defined the same way as for the $^{3}$He
tubes, by applying a 120~keV threshold on the energy deposition of the
charged conversion products in the counting gas. Due
to the complex geometry three different definitions of incident neutrons are
used in this section: incident for the entire detector; incident for a
panel; and incident for a straw.
A neutron is counted as incident for the entire
detector only once, when it enters the wall material of a
tube for the first time.
On the contrary, a neutron can be counted more than once as incident
for a panel but only if it is scattered back from another panel and
enters a tube in the panel of interest again.
Moreover, a neutron is counted as incident for a straw every time when it enters
its copper layer from the outside.

\begin{figure}[!b]  
  \centering
  \begin{subfigure}{0.5\textwidth}
    \centering
    \includegraphics[width=\textwidth]{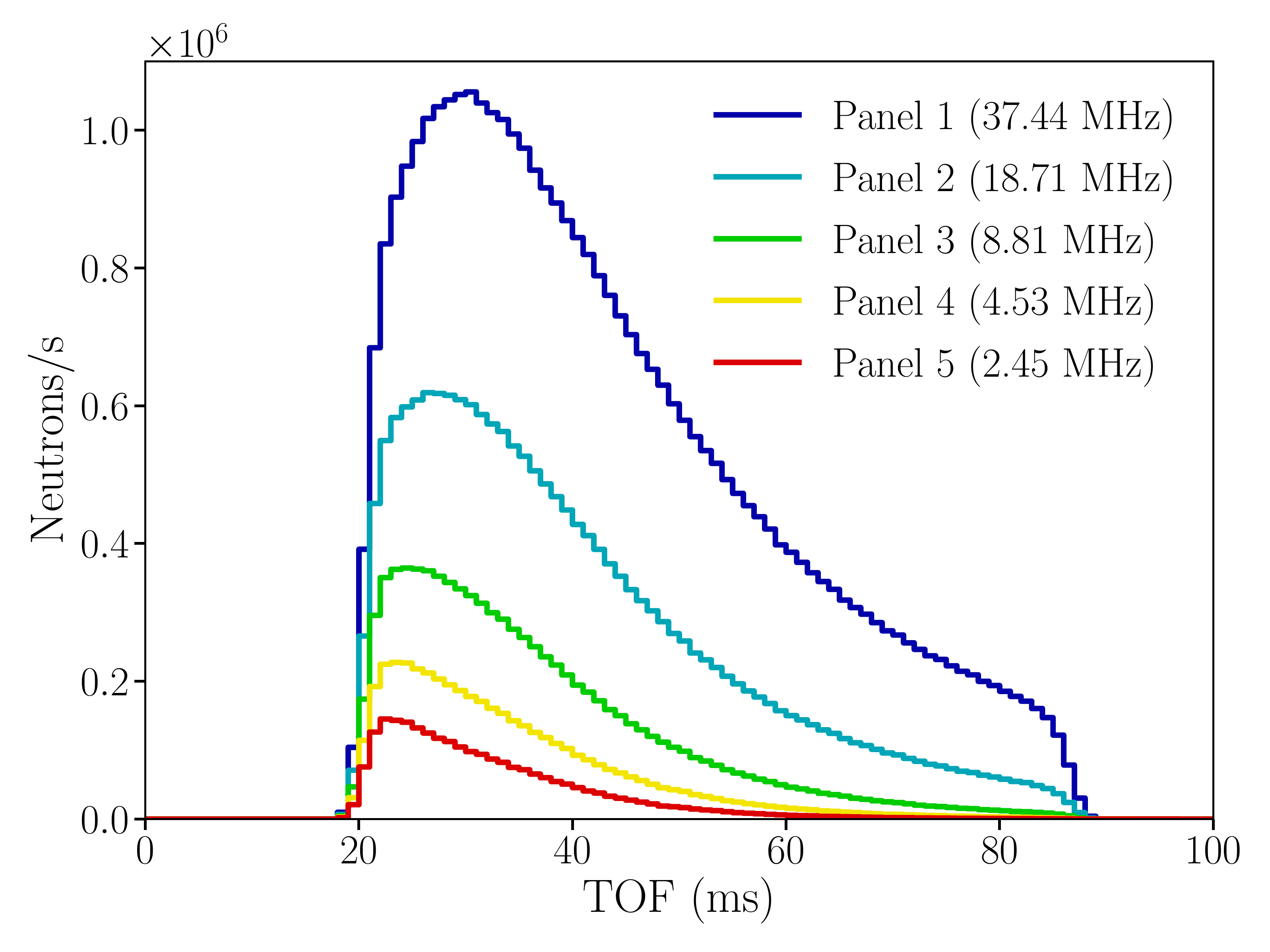}
    \caption{}
    \label{subfigBcsPanelTof}    
  \end{subfigure}%
  \begin{subfigure}{0.5\textwidth}
    \centering
    \includegraphics[width=\textwidth]{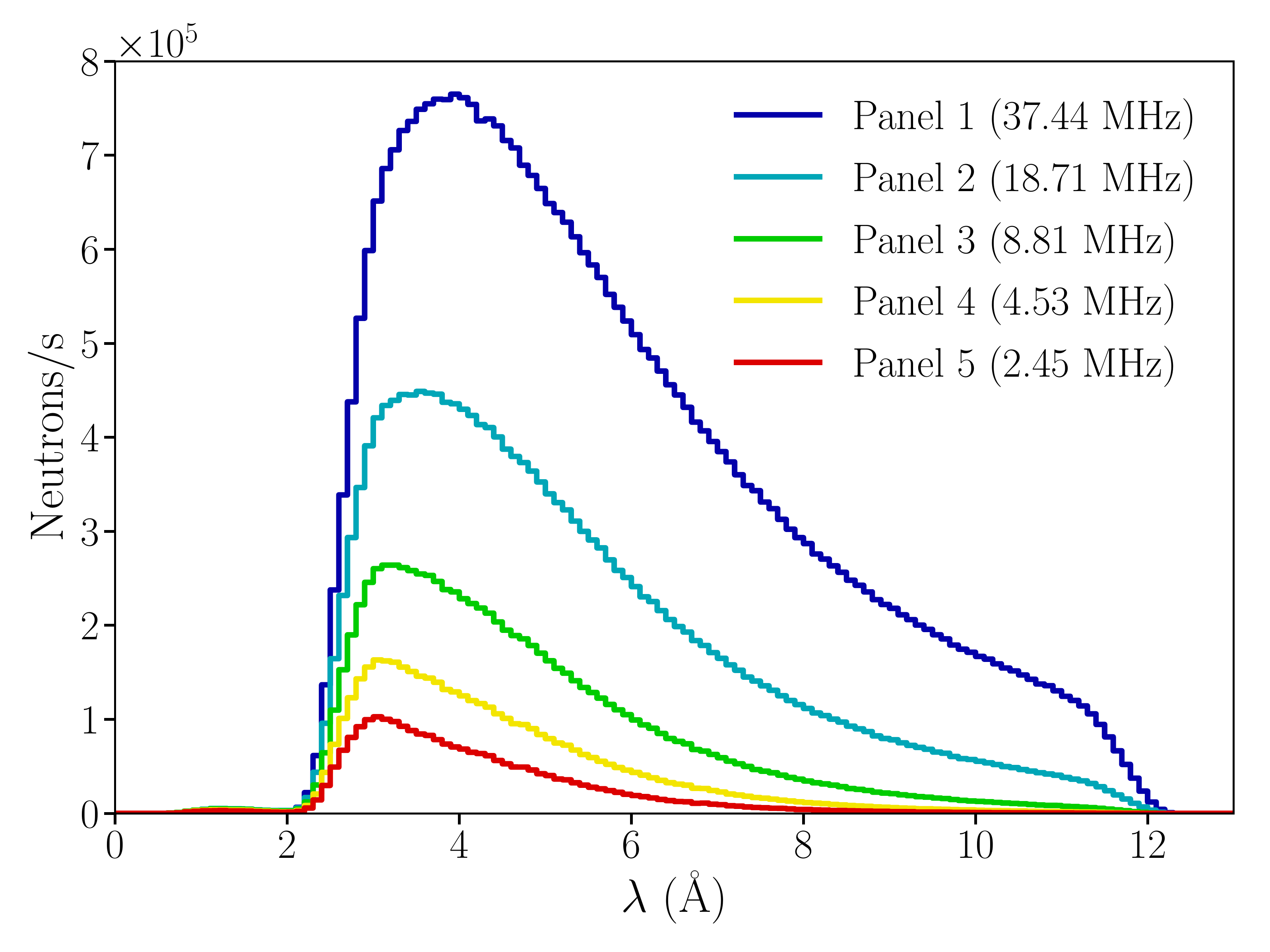}
    \caption{}
    \label{subfigBcsPanelLambda}
  \end{subfigure}
  \caption{\footnotesize Incident TOF (a) and $\lambda$ (b) distributions per
    panel in depth for instrument configuration 1. The respective
    incident rates appear in the legends.}
  \label{figBcsPanelTofLambda}
\end{figure}

\begin{figure}[!h]  
  \centering
  \begin{subfigure}{0.5\textwidth}
    \centering
    \includegraphics[width=\textwidth]{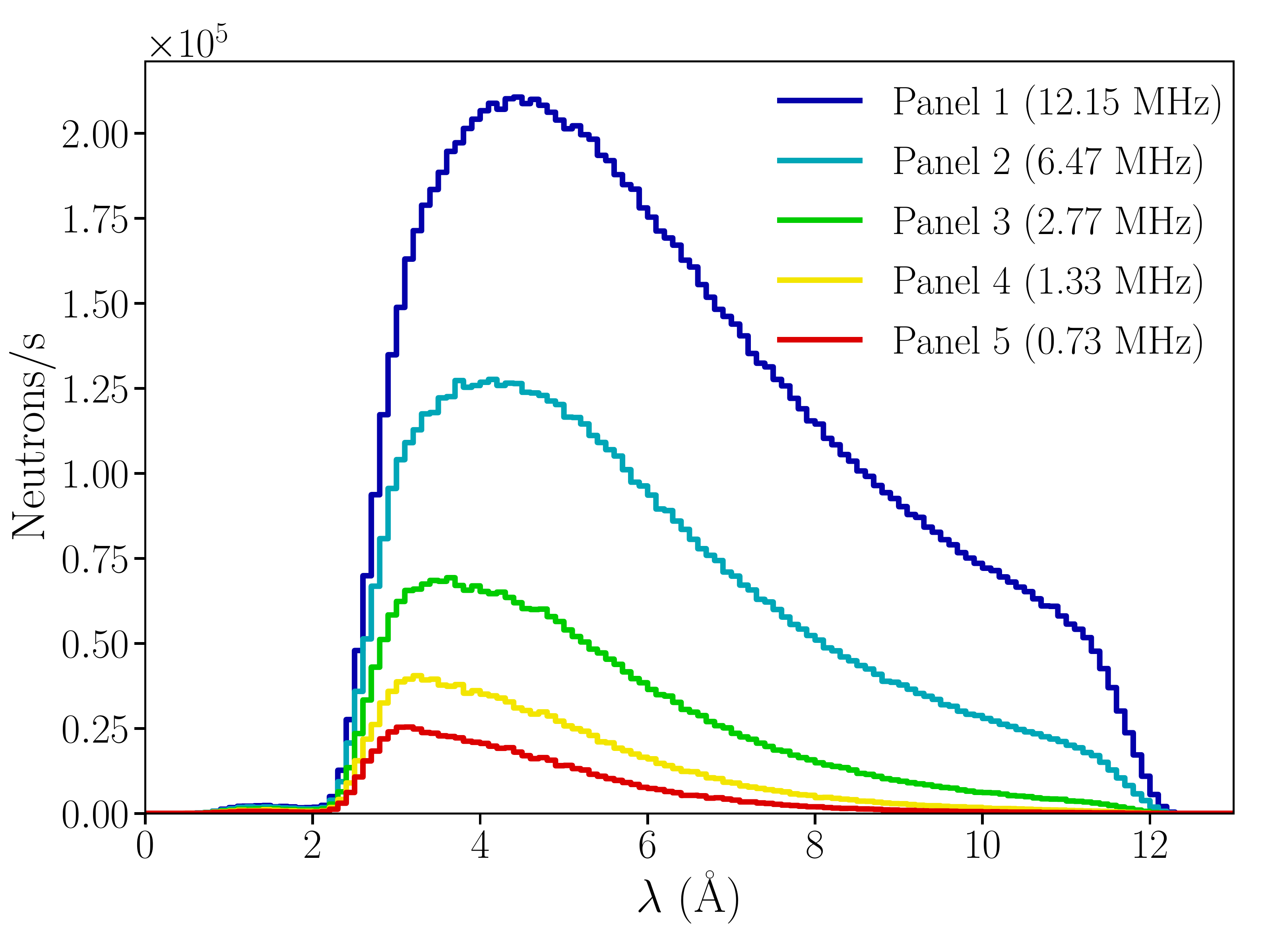}
    \caption{}
    \label{subfigBcsPanelLambdaHit}    
  \end{subfigure}%
  \begin{subfigure}{0.5\textwidth}
    \centering
    \includegraphics[width=\textwidth]{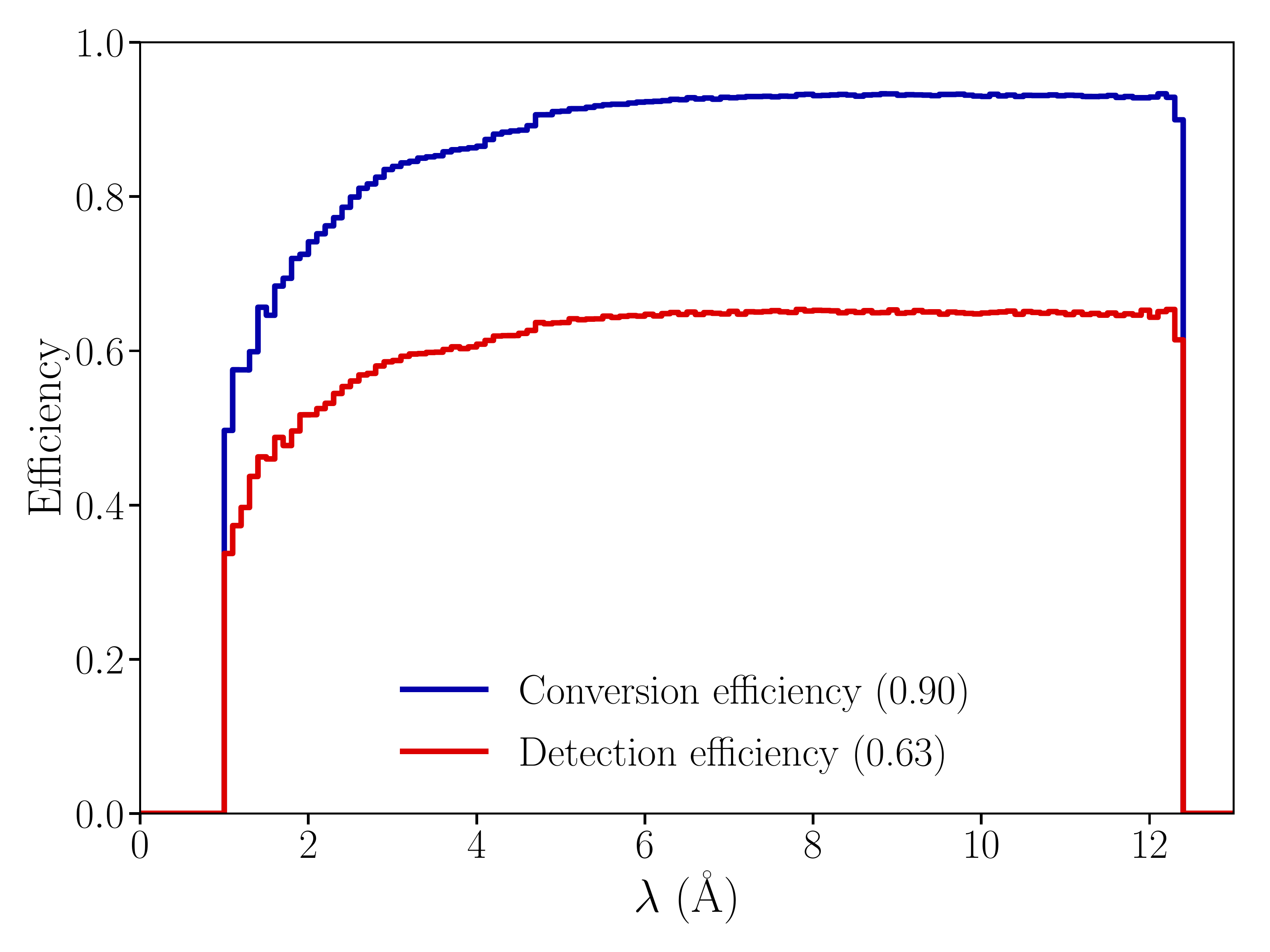}
    \caption{}
    \label{subfigBcsPanelHitEff}
  \end{subfigure}
  \caption{\footnotesize $\lambda$ distribution of detected neutrons per
    panel (a). $\lambda$ dependent neutron conversion and detection efficiency for the entire
    detector (b) for instrument configuration 1. The conversion/detection efficiency is the ratio
    of the total number of converted/detected neutrons in all straws over the number of incident
    neutrons for the entire detector for a particular incident $\lambda$.
    The detection rates per panel and the global average conversion and
    detection efficiencies appear in the legends.}
  \label{figBcsPanelHitEff}
\end{figure}

The incident neutron TOF and $\lambda$ spectra per detector panel
are depicted in Fig.~\ref{subfigBcsPanelTof} and Fig.~\ref{subfigBcsPanelLambda}. 
This approach is important for detectors, which provide depth
information, as every detector layer is exposed to a different neutron
distribution (see Fig.~\ref{subfigBcsPanelLambdaHit}).
From the conversion and detection efficiency of the entire
detector for different wavelengths, depicted in
Fig.~\ref{subfigBcsPanelHitEff}, it is concluded that the efficiency for lower wavelengths is lower -- as expected
-- and 
the global detection efficiency is 63\% with a detection to conversion
ratio of 70\%.
\begin{figure}[bh]
  \centering
    \includegraphics[width=\textwidth]{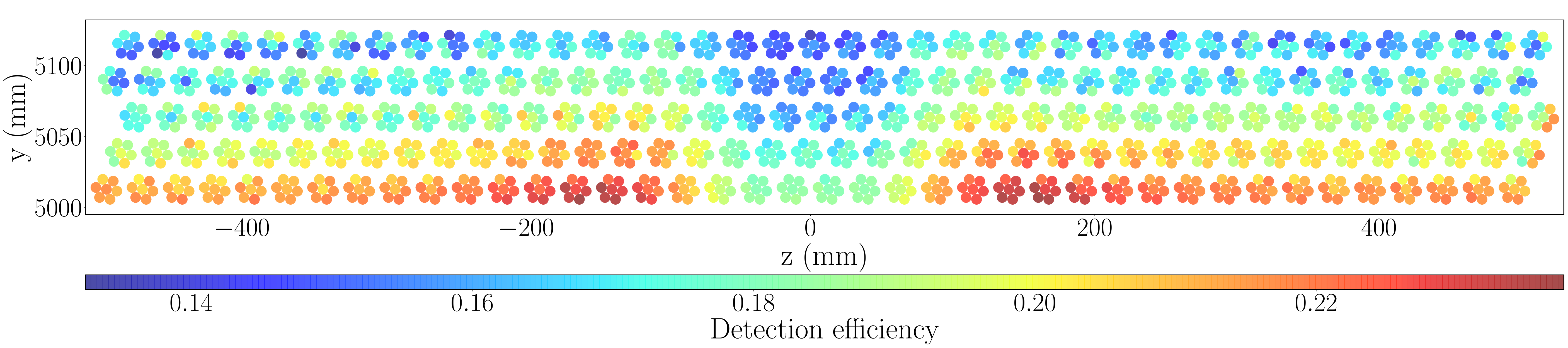}
    \caption{\footnotesize Detection efficiency per straw for instrument
      configuration 1.}
    \label{figBcsStrawEfficiency}
\end{figure}

To get a deeper understanding about the detector
system, the detection efficiency of each straw is shown in
Fig.~\ref{figBcsStrawEfficiency}. It is defined as the number of
detected neutrons in each straw divided by the number of the respective incident neutrons.
The efficiency of each straw is represented with a two-dimensional cross-sectional image
of the detector geometry similar to Fig.~\ref{bcs_zoom}.
The efficiencies of the straws in the first panel indicate that there is
a correlation between the scattering angle and the wavelength and it
is clear that the straws in the front or with clear sight to
the sample have much higher efficiencies than the ones behind them.
All straws are identical and so is their detection efficiency for a particular
neutron wavelength, so the change of the detection efficiency in depth is the
result of the hardening of the neutron spectrum due to the higher
absorption cross-section of neutrons with higher wavelength and the
thermalisation of the neutrons via scattering. This effect in the
spectrum is clearly visible in Fig.~\ref{figBcsPanelTofLambda}.

\begin{figure}[!h] 
  \centering
  \begin{subfigure}{1.0\textwidth}
    \centering
    \includegraphics[width=\textwidth]{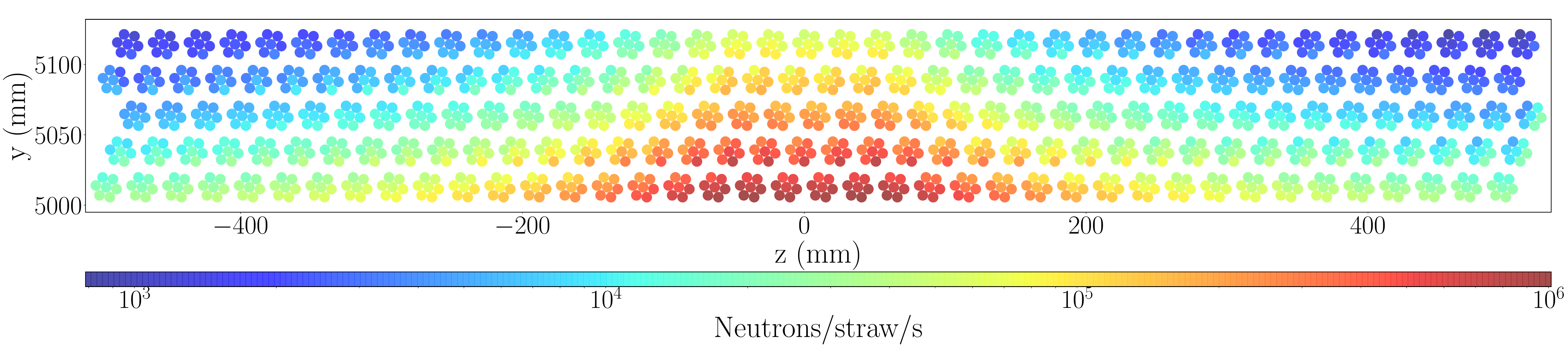}
    \caption{\footnotesize Time-averaged incident rate for per BCS straw.}
    \label{subfigBcsFullIncRate}
  \end{subfigure}
  \centering
  \begin{subfigure}{1.0\textwidth}
    \centering
    \includegraphics[width=\textwidth]{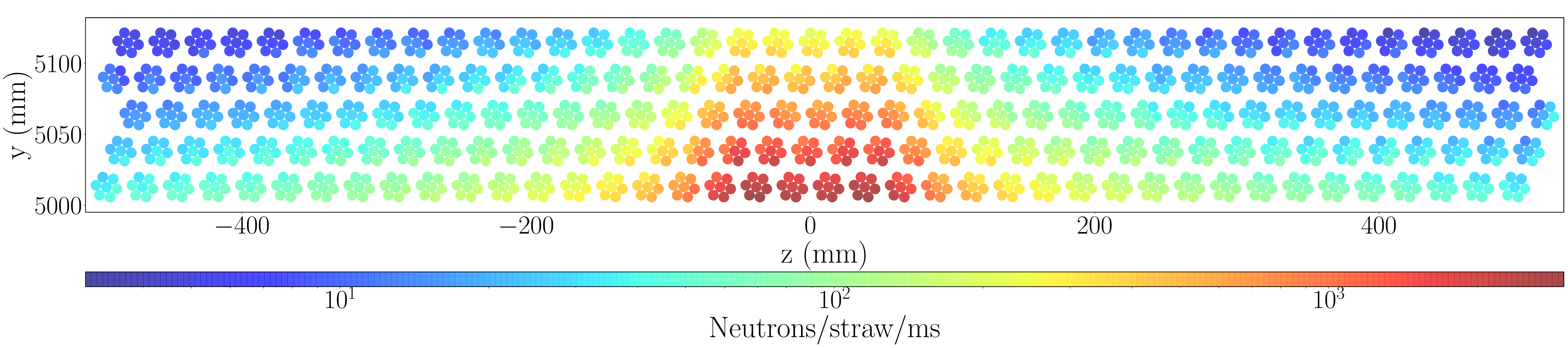}
    \caption{\footnotesize Peak incident rate per BCS straw.}
    \label{subfigBcsFullHitRateNEW}
  \end{subfigure}%
  \caption{\footnotesize Time-averaged and peak incident rate per BCS straw for instrument configuration 1.}
  \label{figBcsIncHitRate}
\end{figure}

Fig.~\ref{figBcsIncHitRate} depicts the time-averaged and peak
incident rate for each straw represented the same way as the detection
efficiency of the straws. The peak rates are extracted the same way as for the $^{3}$He
tubes, by counting neutrons only from a selected 1 ms TOF slice that
results in the highest instantaneous rates. The tubes with the highest
peak incident rates are localised in the center of the panels
because the highest incident rates occur when the relatively high energy
neutrons are scattered on the sample in small angles and most of them enter the same
straws. Fig.~\ref{figBcsIncHitRate} gives an enlarged view of
these straws and also shows the peak detected rates of the straws with the highest
values.
The highest peak rates appear in the straws of the first panel for all three configurations. A summary of the
estimated incident and detection rates can be found in Tab.~\ref{bcs_rate_table}.

\begin{figure}[!h] 
  \begin{subfigure}{0.5\textwidth}
    \centering
    \includegraphics[width=\textwidth]{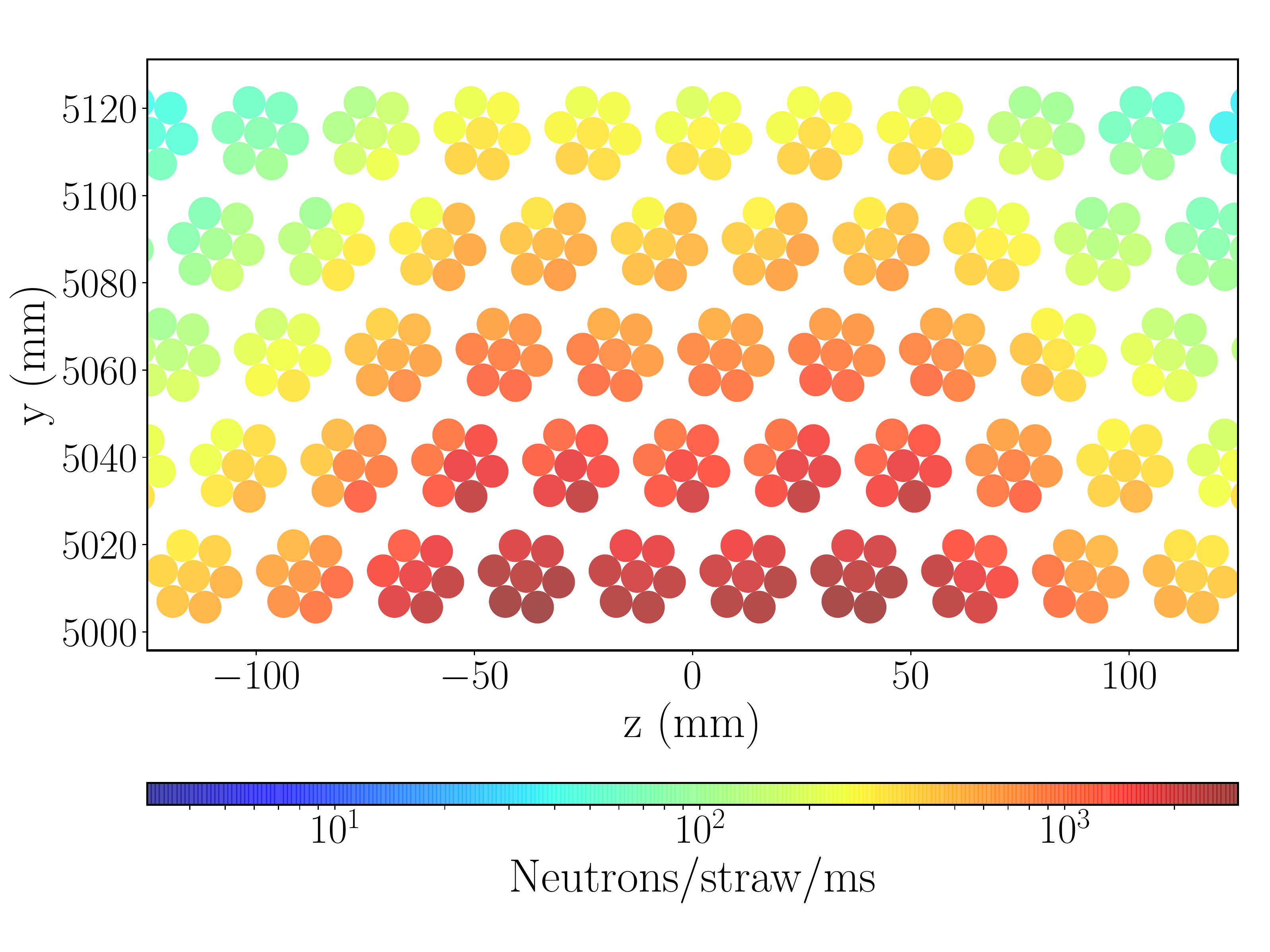}
    \caption{\footnotesize}
    \label{tof2}    
  \end{subfigure}%
  \begin{subfigure}{0.5\textwidth}
    \centering
    \includegraphics[width=\textwidth]{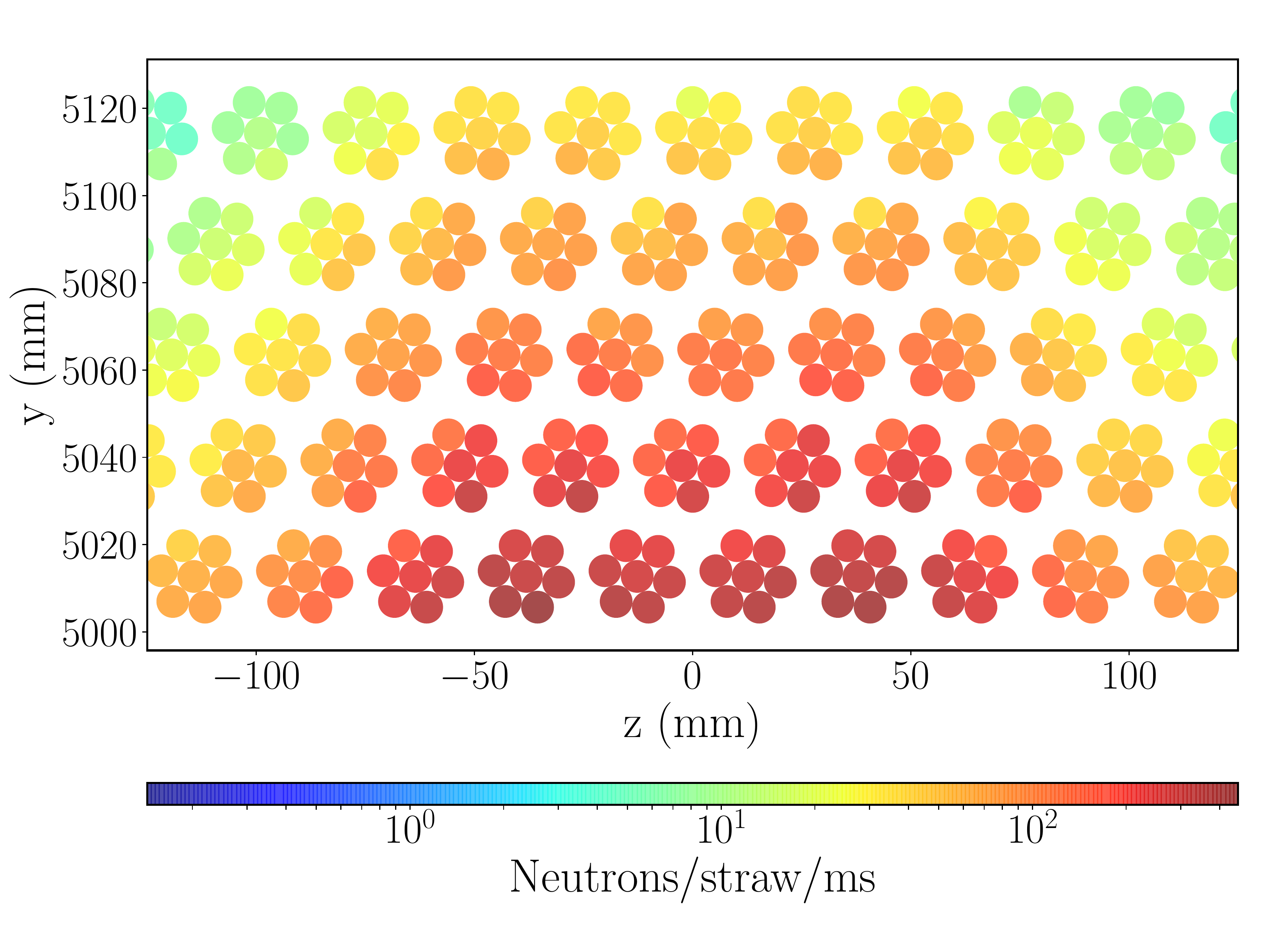}
    \caption{\footnotesize}
    \label{lambda2}
  \end{subfigure}
  \caption{\footnotesize Peak incident (a) and detection (b) rates for the
    central straws for instrument configuration 1.}
  \label{figBcsToF2}
\end{figure}

\begin{table}[!h]
  \centering  
  \caption{\footnotesize Summary table with global and local incident
    and detection rates, both time-averaged and peak ones, for the BCS detector.}
  \begin{tabular}{|c|c|c|c|c|}
    \hline
    config & global average   & global average    & local average                  & local average  \\
           & \textbf{incident} rate & \textbf{detection} rate & \textbf{incident} rate/straw & \textbf{detection} rate/straw \\
    \hline
    1 & 37.3~MHz & 23.5~MHz & 1.01~MHz & 187~kHz \\ \hline
    2 & 7.48~MHz & 4.70~MHz & 211~kHz & 39.0~kHz\\ \hline
    3 & 2.91~MHz & 1.82~MHz & 86.3~kHz & 15.0~kHz   \\ \hline
    \hline
    config & global peak   & global peak    & local peak                  & local peak  \\
           & \textbf{incident} rate & \textbf{detection} rate & \textbf{incident} rate/straw & \textbf{detection} rate/straw \\
    \hline
    1 & 74.3~MHz & 64.2~MHz & 2.99~MHz & 458~kHz \\ \hline
    2 & 15.1~MHz & 13.1~MHz & 651~kHz & 100~kHz \\ \hline
    3 & 6.11~MHz & 5.29~MHz & 283~kHz & 47.0~kHz \\ \hline
  \end{tabular} 
  \label{bcs_rate_table}
\end{table}

Similarly to the $^3$He tubes, the BCS straws are expected to start
saturating at 50-100~kHz. For instrument configuration 1 with the shortest
instrument collimation, peak rates as high as 458~kHz for a single
straw are derived. This implies that the operation of such a detector 
is subject to the same limitations as a $^3$He tube.



\FloatBarrier

\section{Tackling the ESS flux with pixel geometries}

The limitations the tube detector geometries are subject to for the ESS SANS
rates can be alleviated with the use of detectors whose anodes consist
of 2D pixels with individual readout, which leads to
a more efficient distribution of the incoming neutron flux. The pixels
may vary in shape to accommodate an even spread of detection events, 
satisfy spatial resolution requirements, as well as to
geometrically facilitate the polar angle coverage. 

In the following sections two different implementations of 2D pixels
are presented, a detector with square pixels (Solid state Neutron
Detector (SoNDe)) and one with trapezoidal pixels (Boron Array Neutron
Detector (BAND-GEM)). These detectors are developed in the context of
SANS for ESS and are the baseline choices for the respective
instruments. As these detector technologies are high rate capable, only the incident rates
are considered in the following evaluation.
\subsection{Rates for the SoNDe detector}

\begin{figure}[!h]
\centering
\includegraphics[scale=0.25]{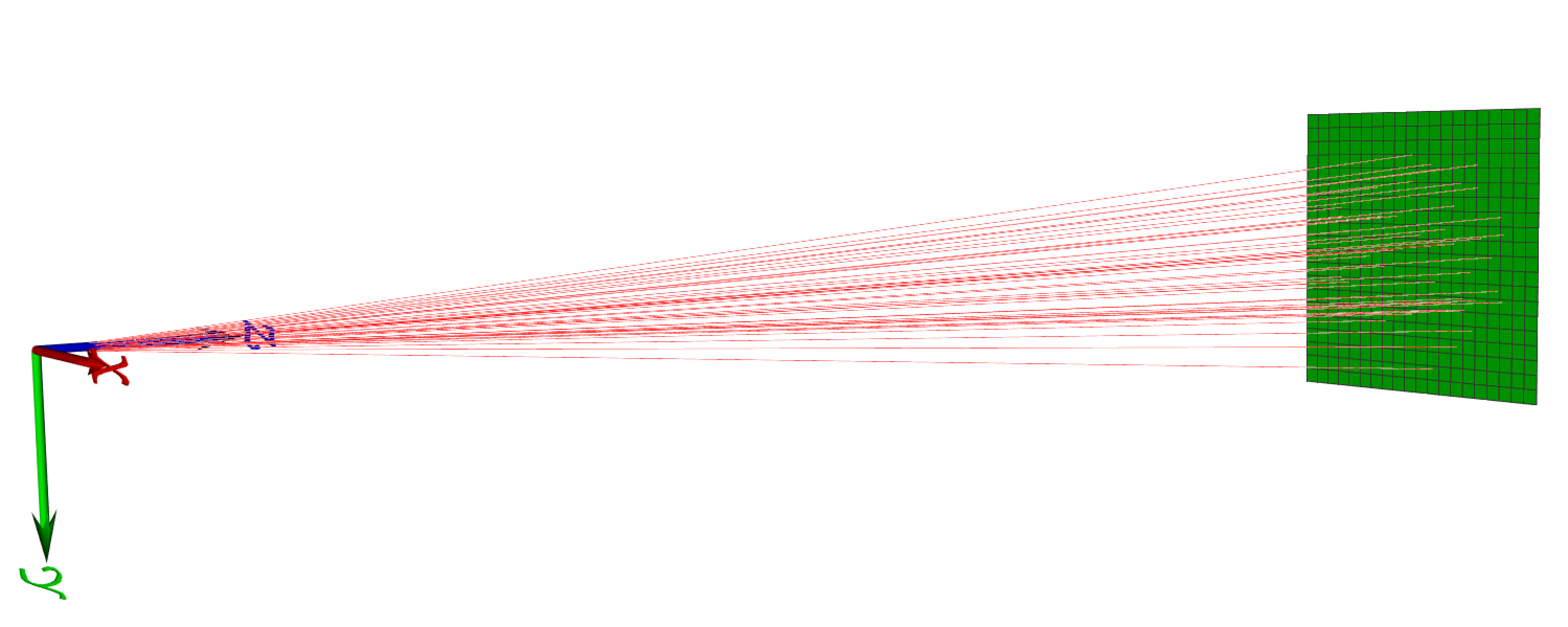}
\caption{\footnotesize Geometry of the SoNDe scintillator tiles in an
  arrangement of 400 (20~$\times$~20) modules occupying a 1~m~$\times$~1~m rear
  detector bank. In red appear primary neutrons originating from the sample.}
\label{sondeGeo}
\end{figure}

The first approach for handling the SANS rates with 2D anode pixels is
the SoNDe detector~\cite{nop_sonde,sonde_patent,sonde_arxiv,sonde_web}, adopted by the SKADI instrument. The
detector surface area is covered by multiple scintillator tiles of
$^6$Li-glass (GS20$\textsuperscript{\textregistered}$~\cite{gs20})
with the respective Multi-anode PMT (MaPMT) placed right behind
them. Each tile is 48.5~mm~$\times$~48.5~cm in size for the chosen MaPMT
model (H12700~\cite{12700}) and is served by 64 (8~$\times$~8) MaPMT pixels. The
detector consists of 400 modules in total, which will be equipping
three different detector banks, as shown in
Fig.~\ref{sonde_drawing}. For the sake of the rate evaluation though,
all 400 modules are assumed to be occupying the rear detector in an
arrangement of 1~m~$\times$~1~m (20~$\times$~20 modules) placed 5~m
after the sample (see Fig.~\ref{sondeGeo}).  

\begin{figure}[!h]  
  \centering
  \begin{subfigure}{0.5\textwidth}
    \centering
    \includegraphics[width=\textwidth]{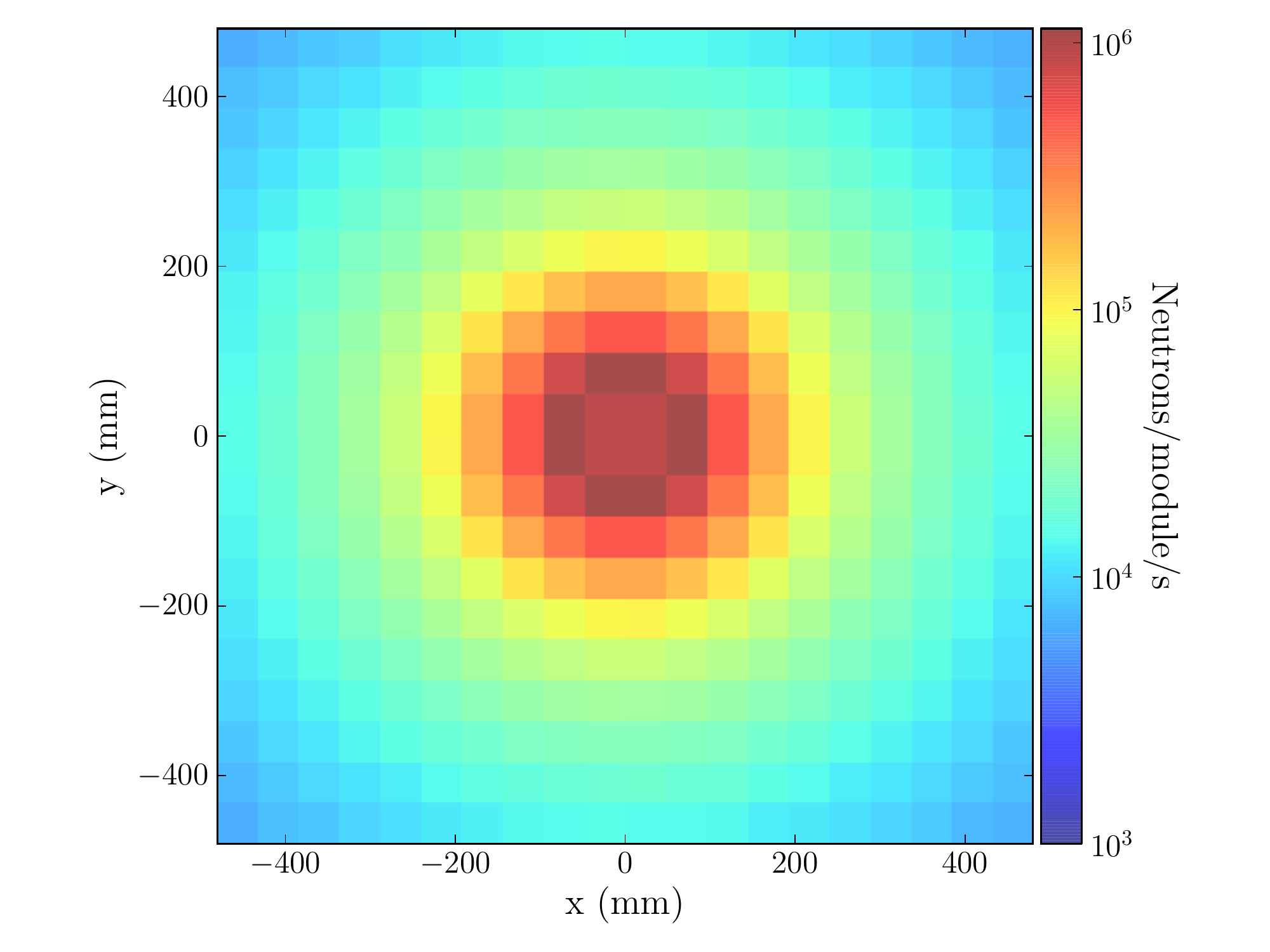}
    \caption{\footnotesize}
    \label{ave_module}    
  \end{subfigure}%
  \begin{subfigure}{0.5\textwidth}
    \centering
    \includegraphics[width=\textwidth]{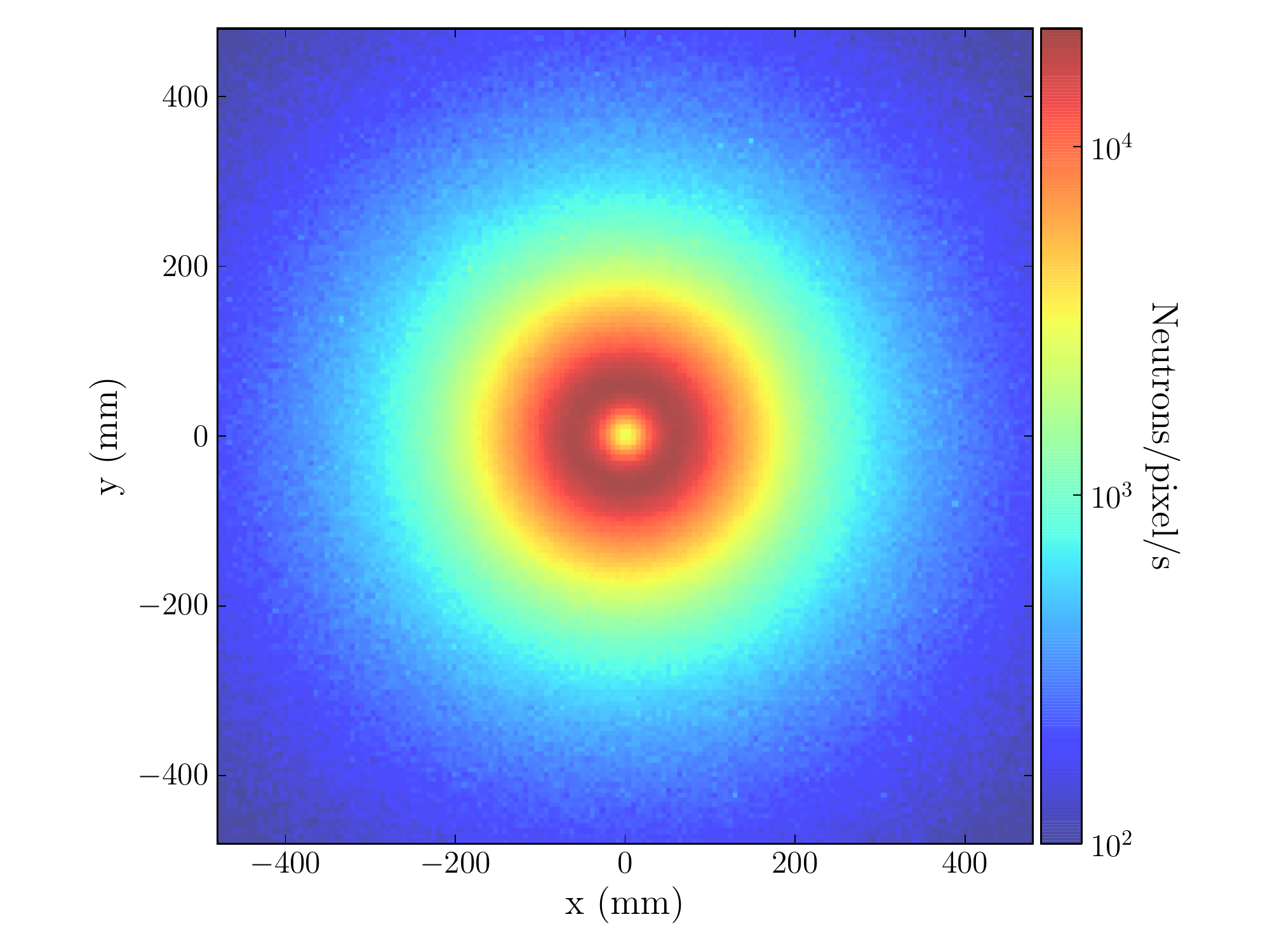}
    \caption{\footnotesize}
    \label{ave_channel}
  \end{subfigure}
  \caption{\footnotesize Time-averaged incident rates per
    SoNDe module (a) or pixel (b) for instrument configuration 1.}
  \label{ave_sonde}
\end{figure}

\begin{figure}[!h]   
  \centering
  \begin{subfigure}{0.5\textwidth}
    \centering
    \includegraphics[width=\textwidth]{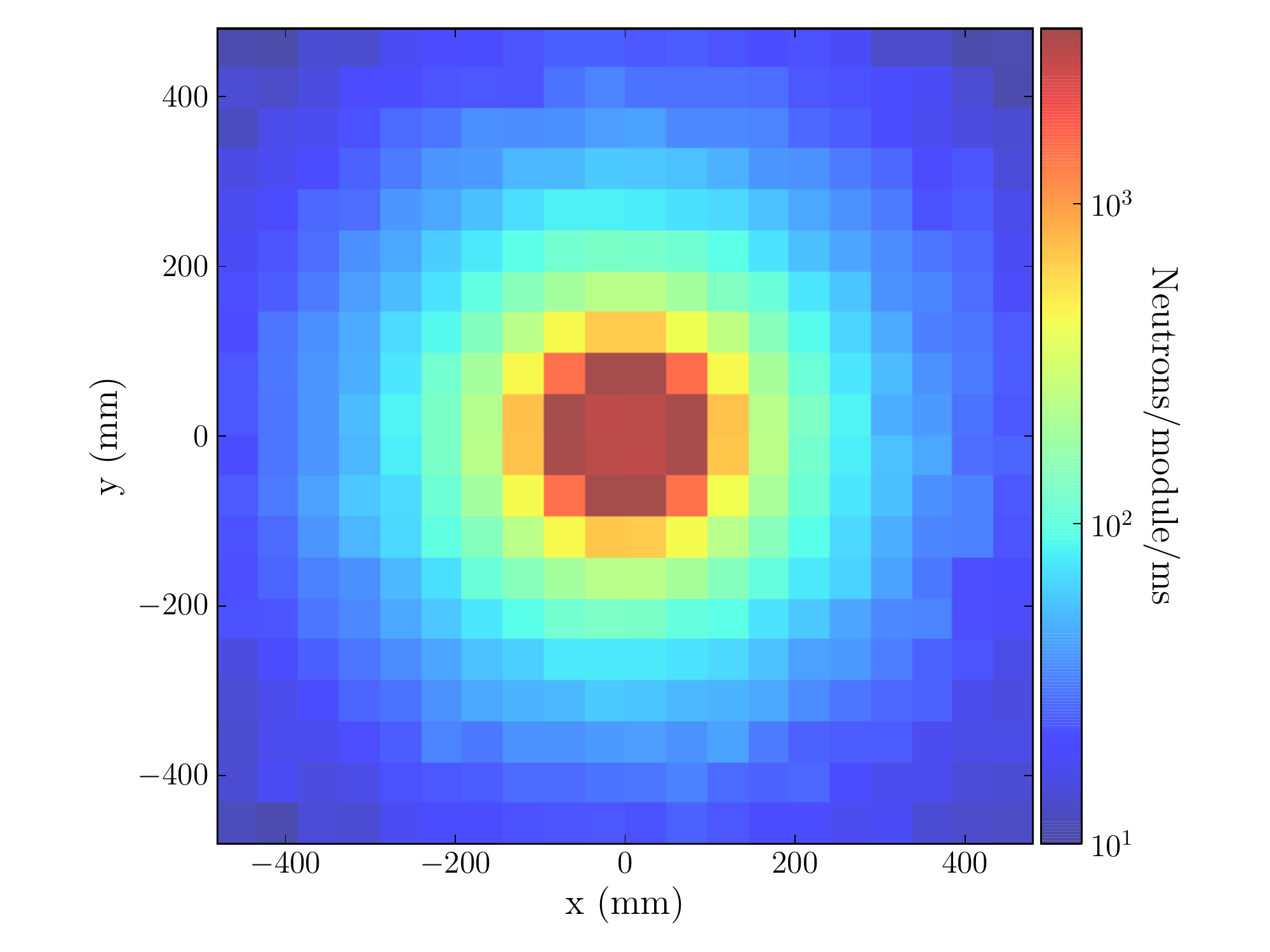}
    \caption{\footnotesize}
    \label{module}    
  \end{subfigure}%
  \begin{subfigure}{0.5\textwidth}
    \centering
    \includegraphics[width=\textwidth]{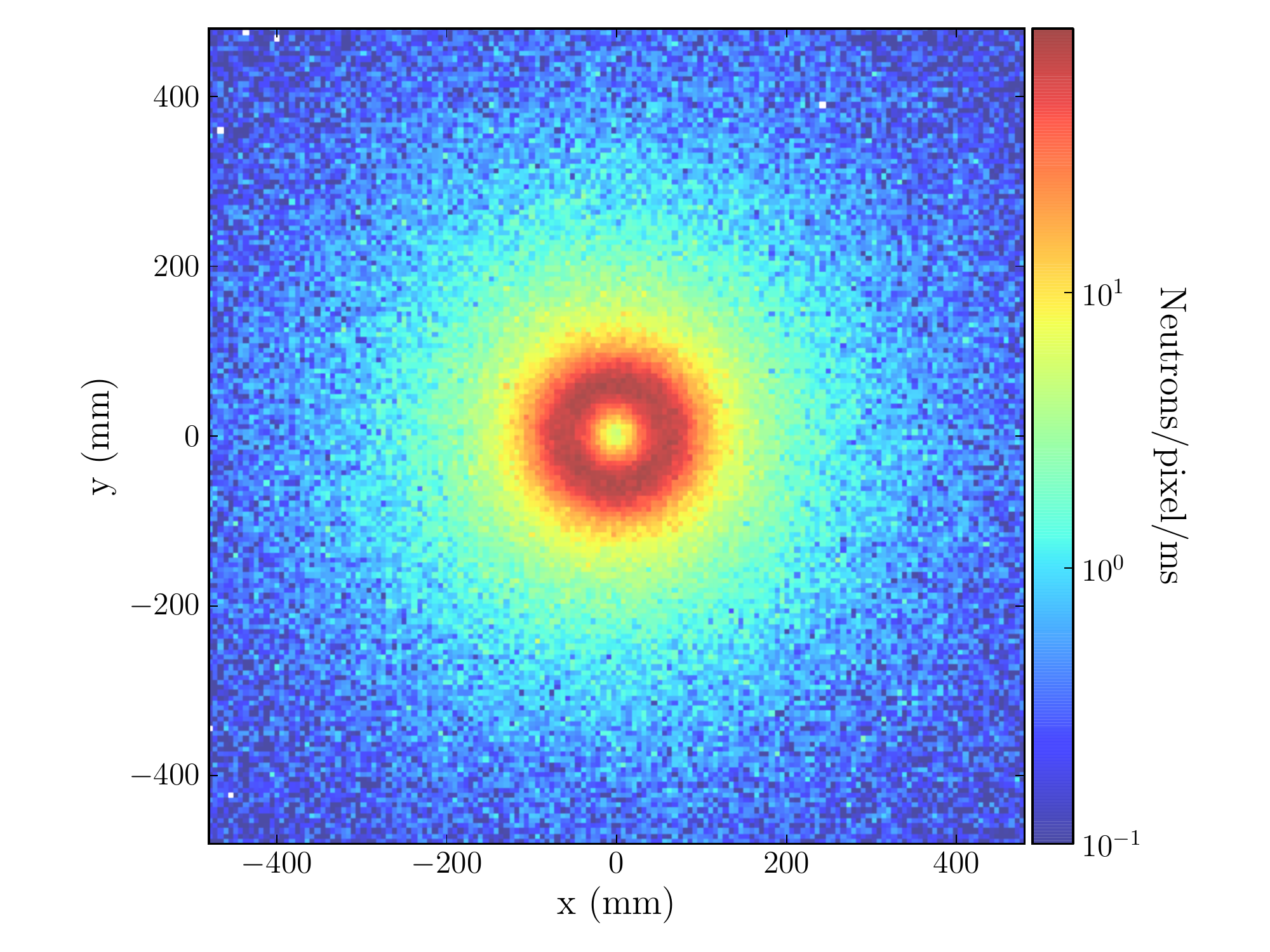}
    \caption{\footnotesize}
    \label{channel}
  \end{subfigure}
  \caption{\footnotesize Peak incident rates per SoNDe module (a) or
    pixel (b)  for instrument configuration 1.}
  \label{peak_sonde}
\end{figure}

An incident neutron is counted the first time it enters the scintillator
glass. The physics taking place inside the scintillator material and beyond
is not included in the Geant4 simulation. Figs.~\ref{ave_sonde} and \ref{peak_sonde} respectively
depict the incident time-averaged and peak rates per SoNDe module and
pixel. 
\begin{table}[!h]
  \centering
  \caption{\footnotesize Summary table with incident time-averaged and
    peak rates for the SoNDe detector.}
  \begin{tabular}{|c|c|c|c|} 
    \hline
    config & global average         & local average                 &    local average      \\ 
           & incident rate & incident rate/module & incident rate/pixel \\  \hline
    1 & 37~MHz   & 1.1~MHz & 21.8~kHz    \\ \hline
    2 & 5.1~MHz & 167~kHz  & 3.6~kHz  \\ \hline
    3 & 2.9~MHz  & 99~kHz   & 2.3~kHz  \\ \hline
    \hline
    config & global peak   & local peak           & local peak             \\   
           & incident rate & incident rate/module & incident rate/pixel  \\    \hline
    1 & 73~MHz & 3.5~MHz & 90~kHz \\ \hline
    2 & 10~MHz & 550~kHz & 21~kHz \\ \hline
    3 & 6~MHz  & 341~kHz & 14~kHz \\ \hline
  \end{tabular} 
  \label{sonde_rate_table}
\end{table}
The total number of pixels for the current Geant4 implementation of
SoNDe is 25600. The SANS technique results in
widespread rates over a large number of pixels, contrary to other
neutron techniques. It does not trigger highly localised
detection events in time and space. Thus, it is imperative for the design of
the data acquisition chain to take into account all aspects of the
rate distribution, and to ensure all buffers are adequate for the expected
data patterns.

It is interesting to look at what these values mean in terms of data
throughput. For the highest flux collimation of the instrument, the
global peak incident rate is about 73~MHz. This means that in the
hottest millisecond of the pulse, 73000 neutrons will enter the
detector. Assuming that every neutron triggers a signal over threshold
on 1-3 pixels, a 64 bit time-stamp
assigned to every fired pixel and an 8-15 bit pixel ID assigned to
pixel location, the total amount of data the detector pushes out
through the optical fibers per second can be calculated as:  
\begin{equation} Data~rate~\big(bps\big) =
  n_{pixels\big/neutron} \times n_{peak~neutrons\big/s} \times \big(
  timestamp + pixelID\big) ~
bits\big/pixel
\label{datadmsc}
\end{equation}
and is in the range of 5-17~Gb/s for a 100\% efficient detector.

The rate values of Tab.~\ref{sonde_rate_table} demonstrate the need
for operating this detector in pixel mode and not in Anger camera
mode, as the shipping of the ADC values required by the algorithm
would make the data rates forbidding for the DAQ. Detectors
are foreseen to run in different operation modes, e.g.\,calibration,
diagnostics, standard. In non-standard operation it is anticipated
that the data rate can significantly increase because of the
additional information collected. However, it is common to tolerate
pile-ups in this context, as the DAQ under
such circumstances usually serves a different purpose and event rejection can be applied. 



\FloatBarrier

\FloatBarrier

\subsection{Rates for the BAND-GEM detector}

\begin{figure}[!h]  
  \centering
  \begin{subfigure}{8cm}
    \includegraphics[width=8cm]{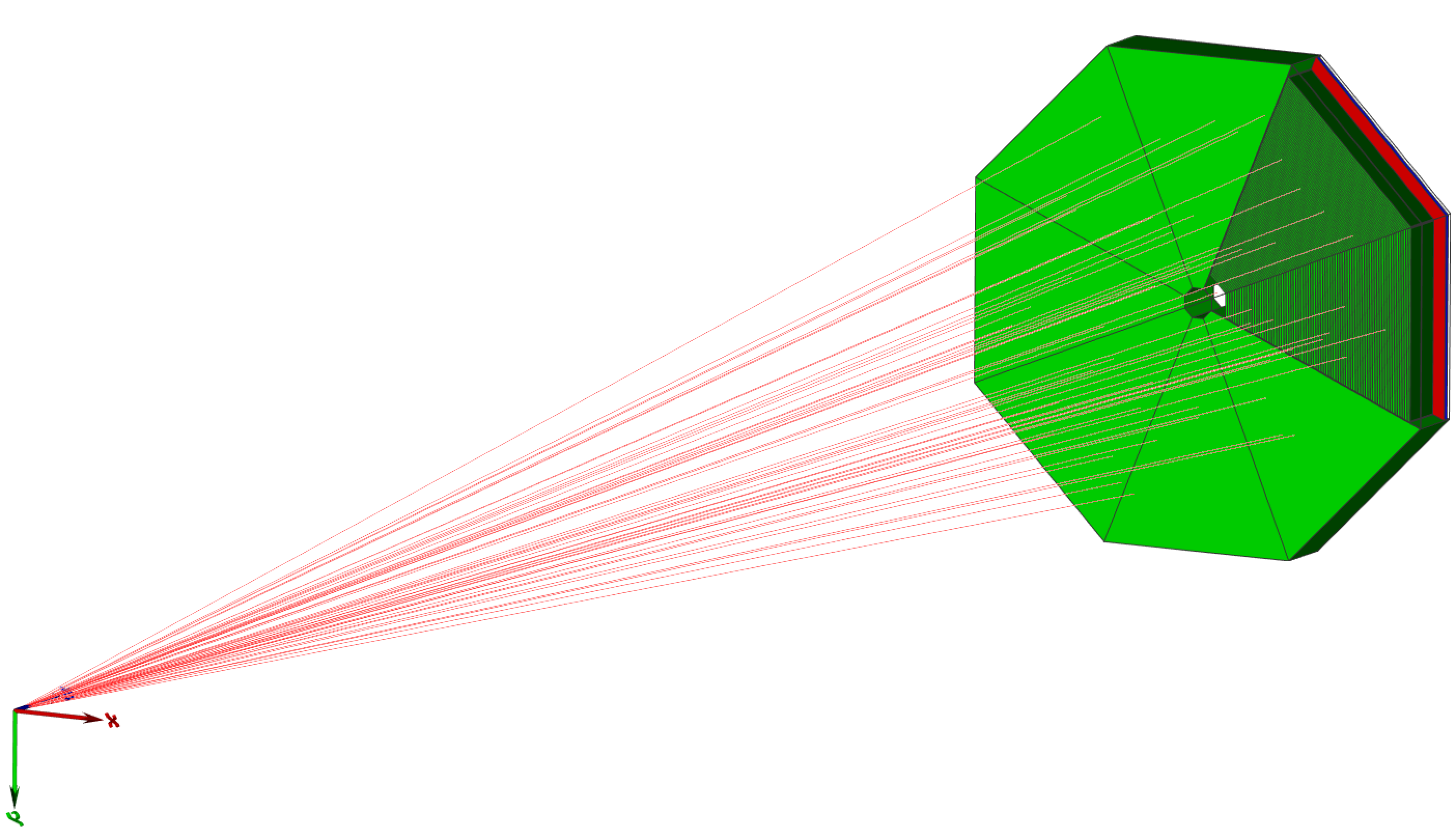}
    \caption{\footnotesize}
    \label{bank1}    
  \end{subfigure}%
  \begin{subfigure}{7cm}
    \centering
    \scalebox{-1}[1]{\includegraphics[width=4cm]{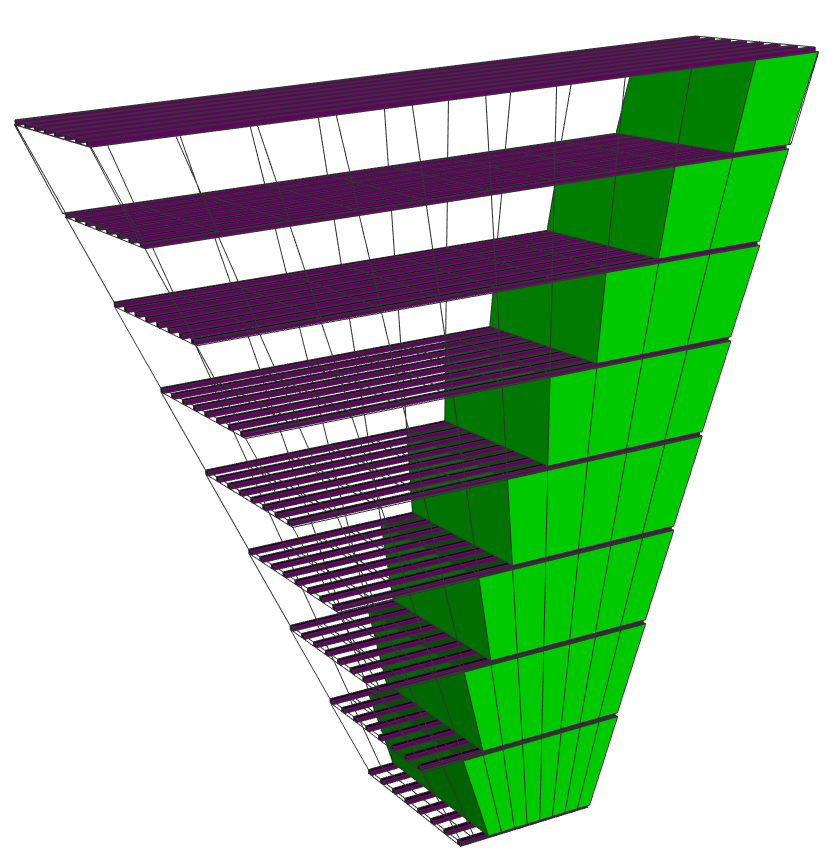}}
    \caption{\footnotesize}
    \label{}
  \end{subfigure}
  \caption{\footnotesize Rear detector bank composed of eight BAND-GEM modules
  arranged in a polygon (a). The  anode panels are placed on the back of the detector (red). The open space in the middle is reserved for
  the installation of a transmission monitor. The cathode structure (purple) of a single segment
  with gas volumes (green) in between (b). The latter geometry shows a
  reduced number of cathodes to allow for better visualisation.}
  \label{geoBG}
\end{figure}

The second detector example of an anode pixel implementation is the
Boron Array Neutron Detector (BAND-GEM)~\cite{bandgem1, bandgem2,bandgem3, bandgem4,giorgiaphd}. It is a trapezoidal GEM module with an intricate
cathode design, which consists of consecutive lamellae forming a
Venetian blind (see Fig.~\ref{geoBG}).  
In order to achieve the polar and azimuthal angle coverage for the
rear detector of a SANS instrument, several trapezoidal BAND-GEM
modules are assembled together to create a polygonal cross section as
in Fig.~\ref{bank1}. 


\begin{figure}[!h]
 \centering
 \includegraphics[scale=0.4]{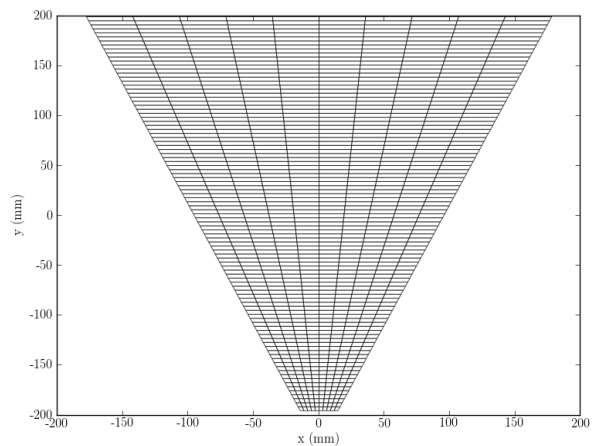}
 \caption{\footnotesize  Anode pad geometry for a BAND-GEM module. In the particular implementation there are 98 rows of pixels with
 10 pixels in each row.} 
 \label{geoBGAnode}
\end{figure}

The anode on which the electrical signal is induced is divided in
equal slices of 10$^{\circ}$ in the azimuthal ($\phi$) direction and
bands with a constant pitch of 4~mm in the polar angle ($\theta$)
direction (see Fig.~\ref{geoBGAnode}). The shape introduced for this
study does not perfectly match the engineering drawing of the actual
demonstrator but the simplification serves the purpose of the rate
analysis without a significant compromise. Practically, the pixels can be
further subdivided in the azimuthal direction at the outer polar
angles to locally increase the spatial resolution or merged at very
low angles, where their size is below the required spatial resolution
to reduce their number.   

\begin{figure}[!h]  
  \centering
  \begin{subfigure}{0.5\textwidth}
    \centering
    \includegraphics[width=\textwidth]{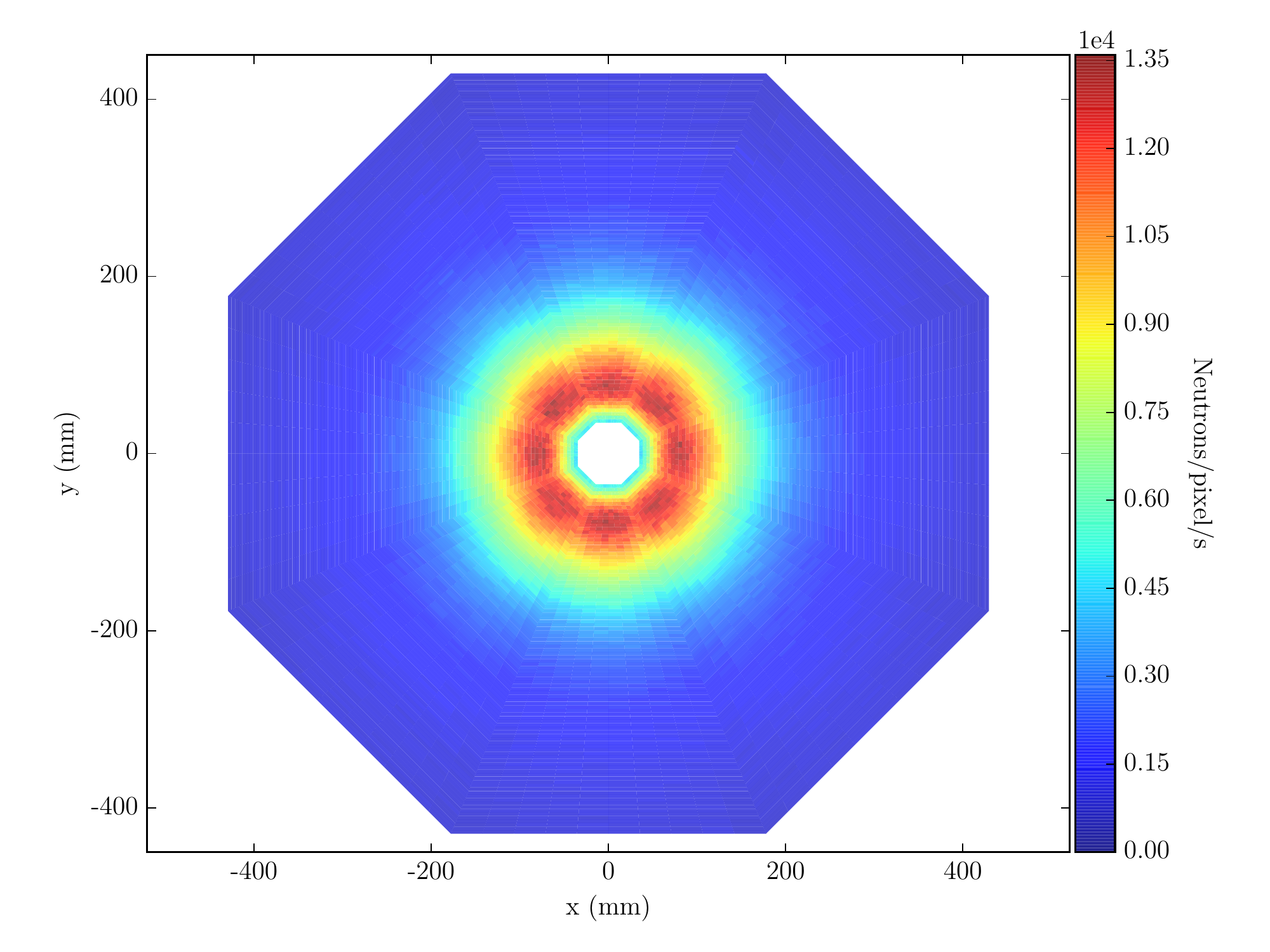}
    \caption{\footnotesize }
    \label{}    
  \end{subfigure}%
  \begin{subfigure}{0.5\textwidth}
    \centering
    \includegraphics[width=\textwidth]{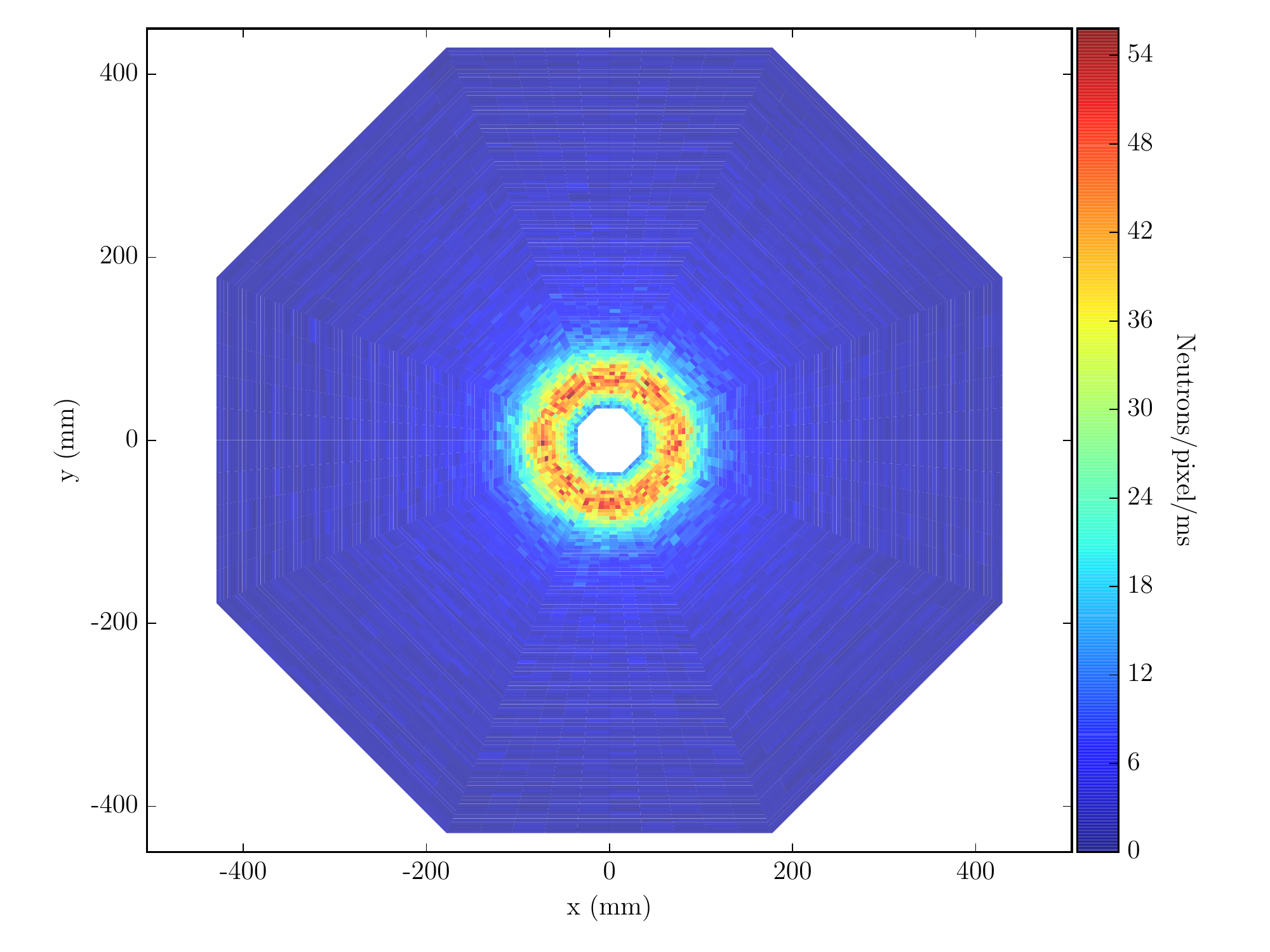}
    \caption{\footnotesize }
    \label{}
  \end{subfigure}
  \caption{\footnotesize Time-averaged (a) and peak (b) incident rates for the
  BAND-GEM detector and instrument configuration 1.}
  \label{bg_average}
\end{figure}

Following the same analysis steps as with the previous detectors, the
derived incident rate estimates are summarised in
Tab.~\ref{bandgem_rate_table}. Fig.~\ref{bg_average} demonstrates the
time-averaged and peak incident rates for instrument configuration 1. The BAND-GEM technology has been
proven to successfully handle thermal neutron rates of up to
40~MHz/cm$^2$~\cite{bandgem1}, which renders it an appropriate choice
for the ESS SANS instruments in terms of rate capability.  

\begin{table}[!h]
  \centering
  \caption{\footnotesize Summary table with incident time-averaged and
  peak rates for the BAND-GEM detector.}
  \begin{tabular}{|c|c|c|c|c|} 
    \hline
    config & global average  & local average  & global peak        & local peak               \\
    & incident rate & incident rate/pixel & incident rate & incident rate/pixel\\
    \hline
    1 & 34~MHz  & 13.6~kHz & 68~MHz & 55.9~kHz\\ \hline
    2 & 7~MHz   & 3.1~kHz  & 14~MHz &   18.1~kHz\\ \hline
    3 & 2.7~MHz & 1.4~kHz  & 5.7~MHz & 12.7~kHz\\ \hline
  \end{tabular} 
  \label{bandgem_rate_table}
\end{table}

A single BAND-GEM module contains 980 pixels in the current Geant4
implementation. For the full detector of eight segments, the total
number of channels is 7840. Using Eq.~\ref{datadmsc} with the
assumption of 1-3 triggered pixels/neutron, 8-15 bits for the pixel ID
representation and 100\% detection efficiency the data throughput that
8 segments would push out is approximately 5-16~Gb/s.

\FloatBarrier
 

\section{Rates for the transmission detector}

Aside from the rate capability of the SCS, the transmission detectors installed at the ESS SANS instruments need special
attention. They are exposed to a large fraction of the direct beam and
the increased neutron flux offers the possibility for per pulse
normalisation. The new operation environment raises several questions
regarding the required spatial resolution, detection efficiency,
stability and longevity of such a detector.  

To this end, a square geometry of 6~cm~$\times$~6~cm segmented in
1~cm~$\times$~1~cm size pixels is placed 5~m away from the sample
covering polar angles up to 0.4$^{\circ}$. The simulated data input
uses the same sample model as before, only this time the transmitted
neutrons which do not interact with the sample are also propagated
until the transmission monitor. The total number of incident neutrons
which enter the detector's active volume are counted and summarised in
the form of global time-averaged and peak rates in Tab.~\ref{tr_table}
for all instrument configurations.  
\begin{table}[!h]
  \centering
  \caption{\footnotesize Summary table with global time-averaged and
    peak incident rates for a 6~cm~$\times$~6~cm transmission detector.}
  \begin{tabular}{|c|c|c|} 
    \hline
    config & global average incident rate  &  global peak incident rate  \\
    \hline
    1 & 911~MHz & 3.6~GHz   \\ \hline
    2 & 193~MHz & 785~MHz   \\ \hline
    3 & 78~MHz  & 219~MHz  \\ \hline
  \end{tabular} 
  \label{tr_table}
\end{table}

\begin{figure}[!h]  
  \centering
  \begin{subfigure}{0.5\textwidth}
    \centering
    \includegraphics[width=\textwidth]{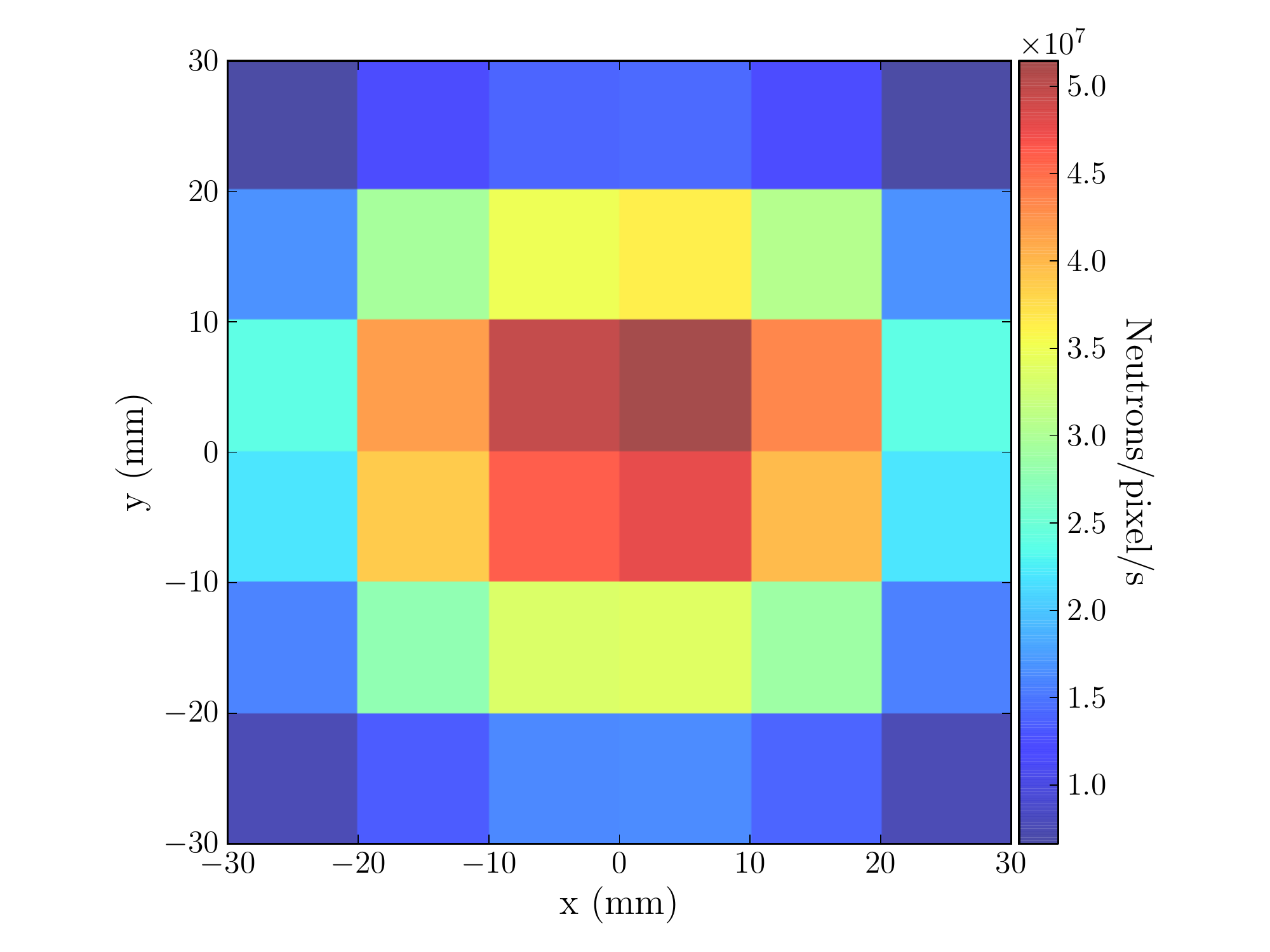}
    \caption{\footnotesize }
    \label{ave_tr}    
  \end{subfigure}%
  \begin{subfigure}{0.5\textwidth}
    \centering
    \includegraphics[width=\textwidth]{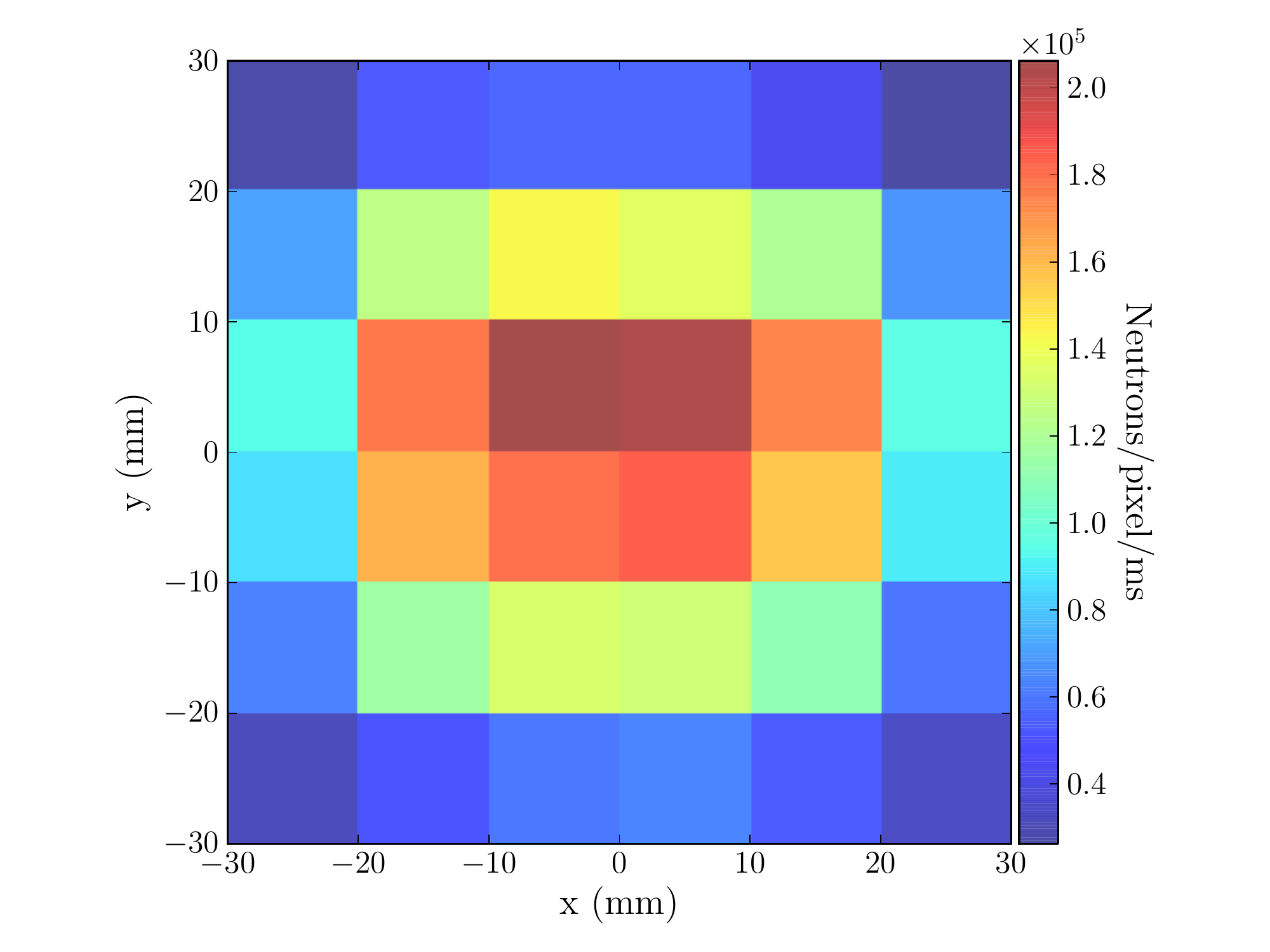}
    \caption{\footnotesize }
    \label{peak_tr}
  \end{subfigure}
   \begin{subfigure}{0.5\textwidth}
    \centering
    \includegraphics[width=\textwidth]{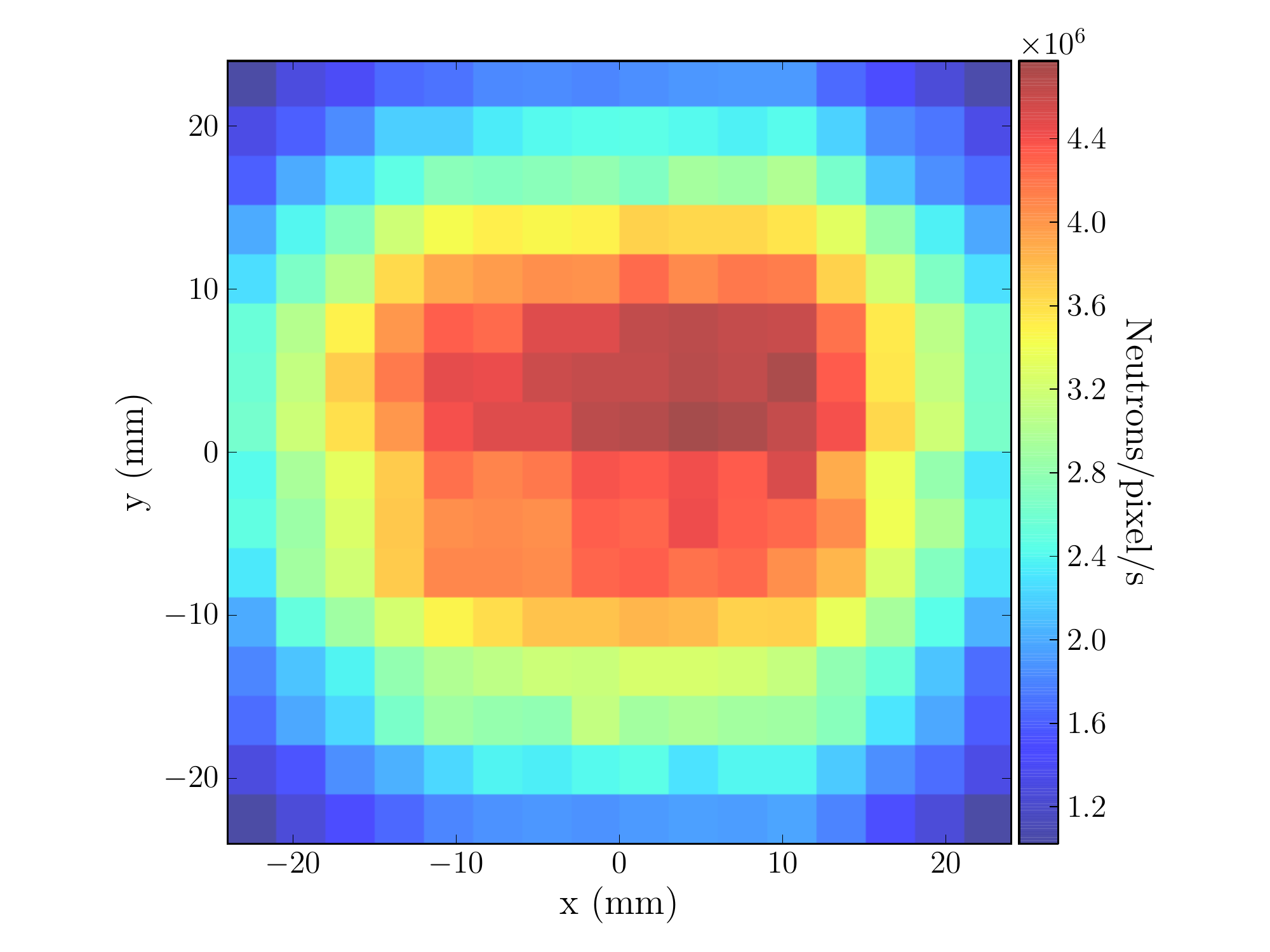}
    \caption{\footnotesize }
    \label{ave_tr_scint}    
  \end{subfigure}%
  \begin{subfigure}{0.5\textwidth}
    \centering
    \includegraphics[width=\textwidth]{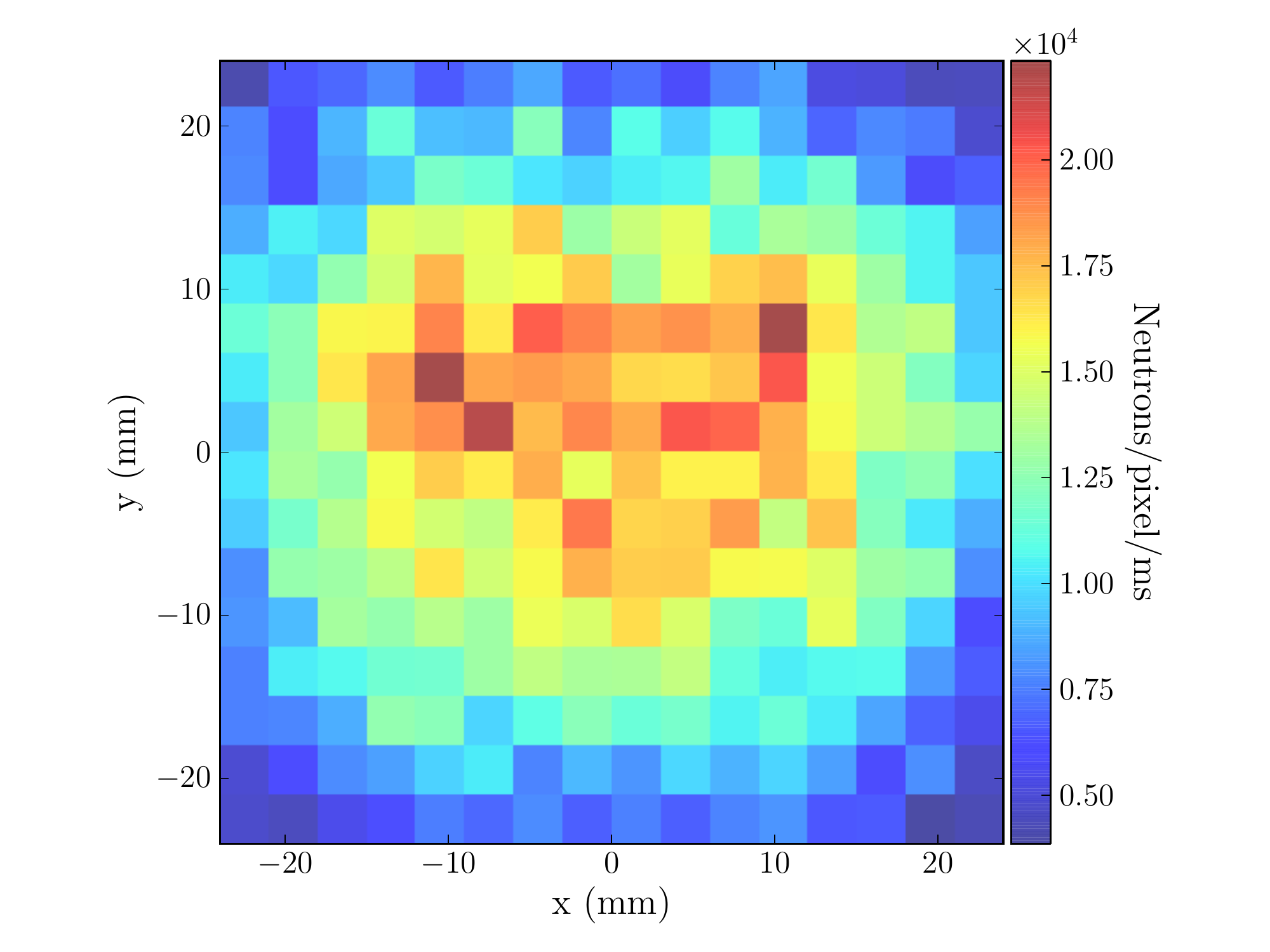}
    \caption{\footnotesize }
    \label{peak_tr_scint}    
  \end{subfigure}
 \begin{subfigure}{0.5\textwidth}
    \centering
    \includegraphics[width=\textwidth]{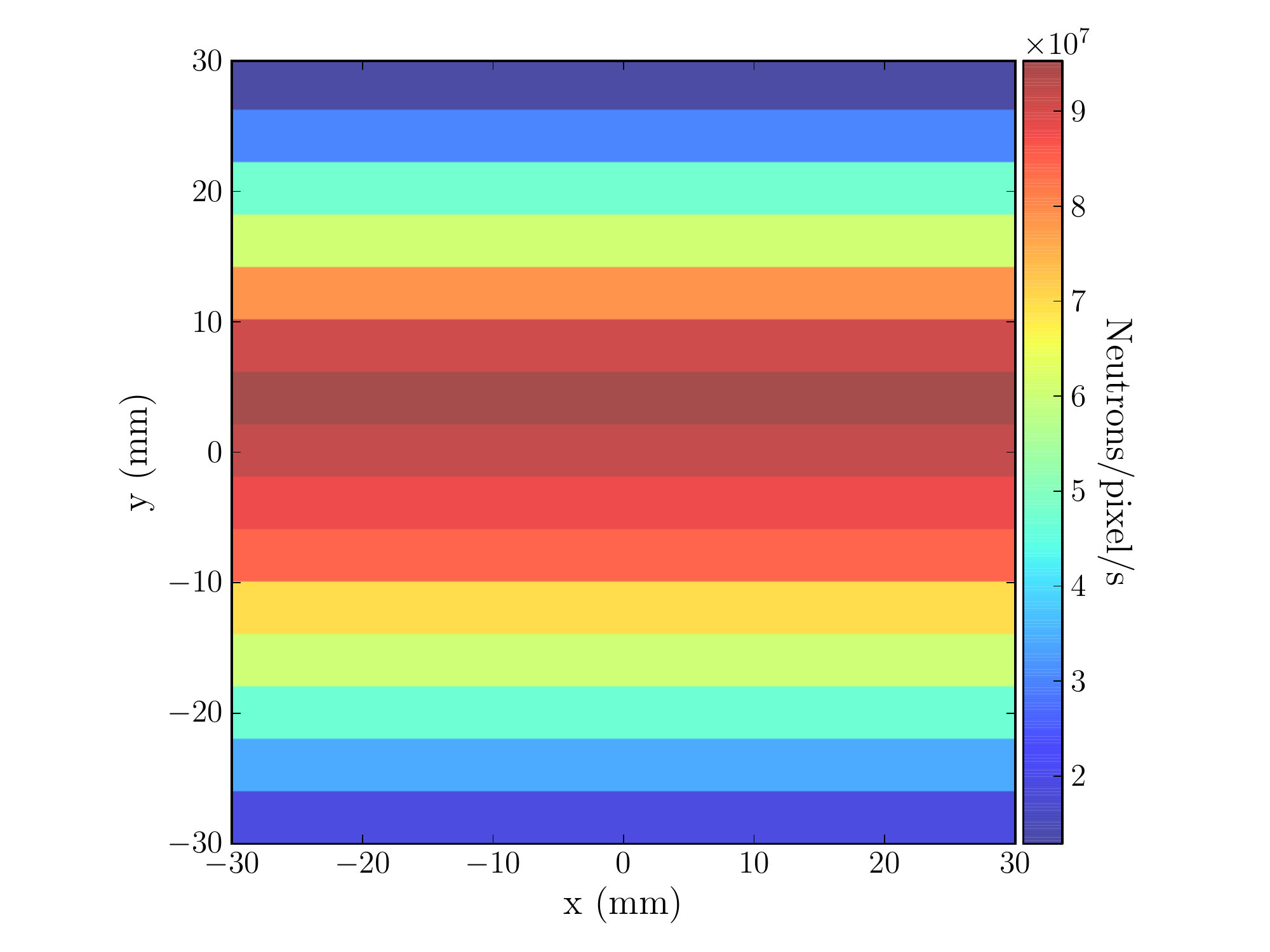}
    \caption{\footnotesize }
    \label{ave_tr_wire}    
  \end{subfigure}%
  \begin{subfigure}{0.5\textwidth}
    \centering
    \includegraphics[width=\textwidth]{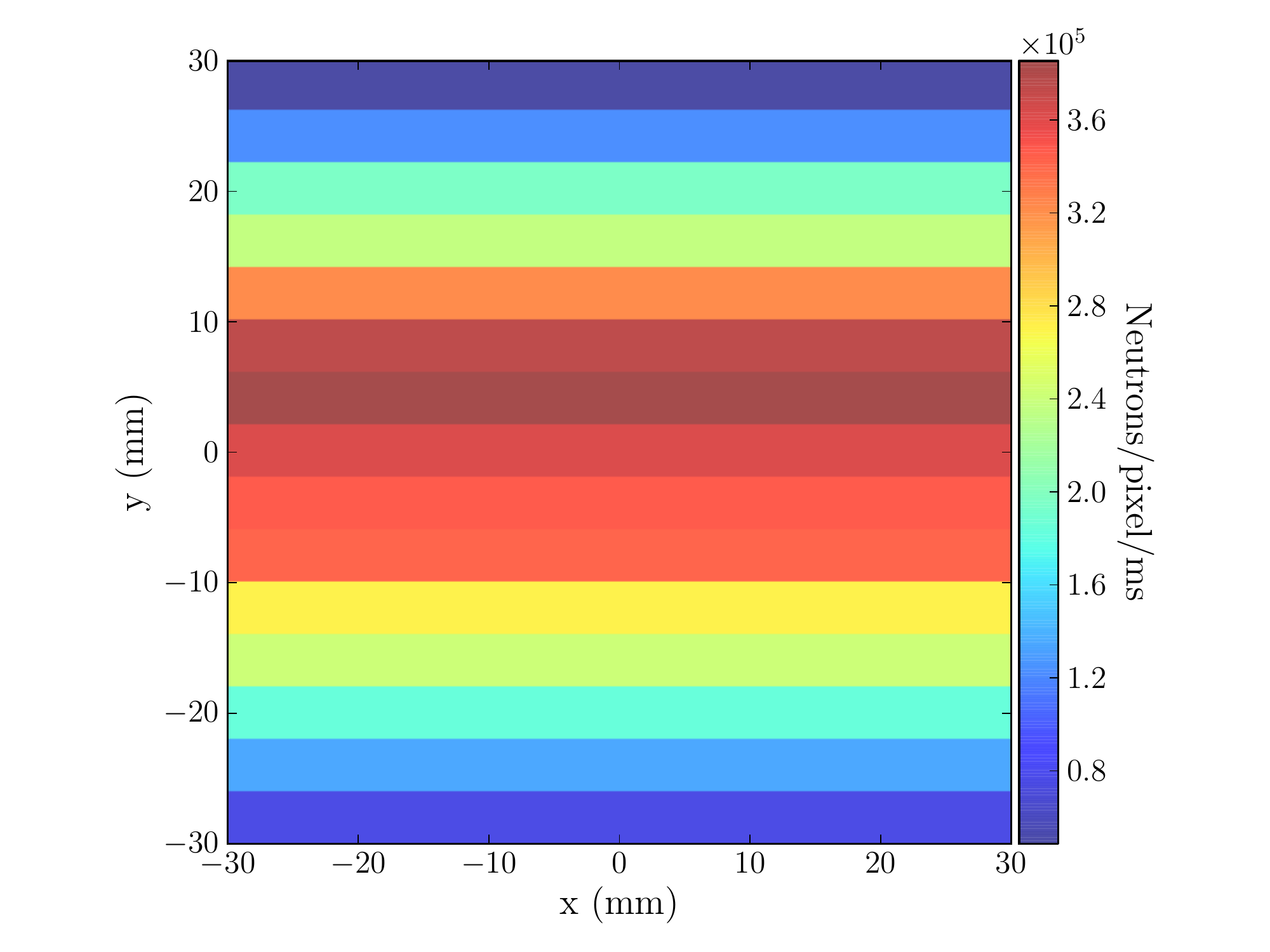}
    \caption{\footnotesize }
    \label{peak_tr_wire}    
  \end{subfigure}
  \caption{\footnotesize Incident time-averaged (left) and peak (right) rates
    for a transmission detector of a 1~cm~$\times$~1~cm pixel size (a,
    b), a
    3~mm~$\times$~3~mm pixel size (c, d) and a wire gas counter (e, f)
    for instrument configuration 1.}
  \label{tran}
\end{figure}

A non-exhaustive list of implementations is evaluated below; a generic pixelated
anode with a 1~cm~$\times$~1~cm pixel size, a SoNDe-like
implementation based on the H9500 MaPMT model~\cite{9500} with a
3~mm~$\times$~3~mm pixel size, a wire gas counter of 15 wires at a
4~mm pitch and an ionisation chamber operated in current mode. The respective local
time-averaged and peak incident rates are presented in Fig.~\ref{tran}
and Tab.~\ref{tr_table_local}.
\begin{table}[!h]
  \centering
  \caption{\footnotesize Summary table with incident time-averaged and
    peak rates for various implementations of the transmission detector for all instrument
    configurations.}
  \begin{tabular}{|c|c|c|} 
    \hline
    config  & local average incident rate  &  local peak incident rate  \\
    & /cm$^2$ /9mm$^2$ /wire & /cm$^2$ /9mm$^2$ /wire\\
    \hline
    1 & 52 / 5 / 95~MHz & 206 / 22 / 385~MHz \\ \hline
    2 & 30 / 4 / 37~MHz & 134 / 19 / 154~MHz \\ \hline
    3 & 17 / 3 / 20~MHz &  49 / 11 /  81~MHz \\ \hline
  \end{tabular} 
  \label{tr_table_local}
\end{table}

For a 1~cm$^2$ pixel anode, a peak incident rate of 200~MHz/cm$^2$
translates to an average pulse spacing of 5~ns and a typical detector
rise time of 500~ps, rendering  a full efficiency implementation
impossible. Assuming a 1\% detection efficiency, the peak
detection rate drops to 2~MHz/cm$^2$. A fast detector with a single
converter layer, e.g.\,a GEM, could satisfy this requirement and
possibly allow for a higher detection efficiency up to 15-20\% before it
saturates. The analysis above also demonstrates that the 1~cm$^2$ pixel size is the
upper limit for use in a high flux environment. 

A SoNDe-like implementation with a smaller square pixel size of 3~mm~$\times$~3~mm, equipped with a fast scintillator like the GS20 glass is
thus another possibility. The upper incident limit estimated here is
at 22~MHz/pixel and given the detector has a high efficiency
(>75\%~\cite{sonde_arxiv}), either the incoming neutron flux has to be
significantly attenuated or the detection efficiency has to be reduced for the
module to operate.

A wire gas counter without position information, covering the same
active area, e.g. with 15 wires at 4~mm pitch, is explored next. With
an incident peak rate of 385~MHz/wire (see Tab.~\ref{tr_table_local})
and a detection efficiency of 10$^{-4}$ (e.g.\,with N$_2$ as counting gas), the
order of magnitude for the peak detection rate is still
around 40~kHz/wire, which poses a challenge for the stability of
such a detector.  
 
Last but not least, a $^3$He- or $^{10}$B-based ionisation chamber
can be considered, if only integrated flux is required from a transmission
detector. Both gas and solid state implementations are used in high
flux environments for reactor or beam monitoring. They are operated in current mode,
i.e.\,the collected charge is integrated in time. The current is
proportional to the direct ionisation charge released in the active volume, as
the chamber is operated with no gain. This type of
detectors have demonstrated operation stability and can perform over a large
dynamic range. However, as they are uncommon in the neutron scattering
field, development is needed to customise them for the particular
use. Similar development steps would be necessary for other more exotic detector solutions,
in order to ensure their performance suitability and
stability~\cite{b10nano, thickGEM, b10suspension}.

As the transmission detector is exposed to a large fraction of the
direct beam, the outgoing data rates could prove challenging. Indeed, with the
assumptions of a 20\% detection efficiency, 1-2 pixels triggered per
event, a 64 bit time-stamp per pixel and an 8-10 bit pixel ID, the output
data rate is 53-107~Gbps for instrument configuration 1. To stay
below the 100~Gbps limit of the optical fiber, the output data rate can be
lowered by further reducing the detector efficiency, the accuracy of the time
stamp or by optimising the data format to contain a single time stamp
per event for all triggered pixels. A detailed analysis of the instrument requirements with
respect to the detector ones can set the limits of the design parameters.




\FloatBarrier

\section{Conclusions}

As the ESS source will subject the detectors to unprecedented neutron
fluxes, it is scientifically imperative that the new detector
requirements are fulfilled. One of the bigger challenges is the rate
capability of detectors used in SANS techniques, whilst satisfying
other requirements. An overview of the
incident and detection rate estimates anticipated for the upcoming
SANS instruments at ESS is presented. The rate capabilities of various
detector technologies are discussed for the rear SCS, as well as for the transmission
detector. Lower and upper limits are respectively set, as only
the coherent signal from the sample is used. The time-averaged SCS rates
need to be corrected upwards by a few \% to include the incoherent
component, while the same correction has to be applied downwards for
the transmission rates. A more detailed look at the time dependence
of the incoherent signal would be necessary to adjust the peak rate
values. The study is based on a baseline LoKI instrument model. SKADI
is foreseen to operate with longer collimation lengths compared to
LoKI, which immediately relaxes the detection rate requirements.  

It becomes apparent that tube detectors can potentially
compromise the instrument performance for high flux configurations. A
standard $^3$He tube would saturate already at 100-200~kW source
power. A BCS straw is expected to saturate at 800-1000~kW source
power. The tube parameters can be adjusted, so as to distribute
the detection events over multiple layers but such an improvement
cannot adequately lift the saturation limit of this
detector type. Moreover, the addition of multiple detection layers enhances
the scattering effects with a negative impact on the spatial
resolution of the detector. A custom readout electronics design can
partially alleviate the saturation but again at the cost of spatial
resolution. Last, both tube technologies explored in this study employ a closed gas system,
which prevents counting gas purification and impurity filtering, and has consequences on the
quality of operation with sustained high rates over the
years. Particularly for BCS, long term operation under such conditions
has not been demonstrated. 

Alternative technologies, like SoNDe and BAND-GEM, with 2D anode
pixels are a more suitable choice to exploit the full ESS source
power. The pixel size and shape can be designed for optimal spatial
resolution and rate capability. In the SoNDe case the rate estimates
indicate that the detector needs to be operated in pixelated mode, as the
Anger camera mode would result in data rates too excessive for an affordable DAQ
system. In the BAND-GEM case the open gas system allows for efficient
monitoring of the gas quality and composition, which contributes to the
stability and longevity of the detector. As the amount of channels and
data output dramatically increase, special consideration is needed for
the choice of bandwidths and buffers throughout the DAQ chain.  

Last but not least a first glance at the rates for the transmission
monitor is attempted. Few different technologies are discussed in this
context. An important conclusion is that for a position sensitive
detector, the pixel size cannot exceed the 1~cm$^2$ for saturation
reasons. Moreover, it turns out that the output data rate can be more
demanding for this detector than for the SCS. An understanding of the spatial resolution, efficiency and
accuracy requirements from the instrument side is vital for the
customisation of the proposed technologies, as well as
the design of the DAQ chain.

\acknowledgments
Richard Hall-Wilton, Kalliopi Kanaki and Thomas Kittelmann would like to acknowledge
support from the EU Horizon2020 BrightnESS grant~(676548). Richard
Hall-Wilton and Kalliopi Kanaki would also like to acknowledge support
from the EU Horizon2020 SoNDe grant~(654124). Mil\'an Klausz would
like to acknowledge the supervision of P\'eter Zagyvai. Computing
resources were provided by the DMSC Computing Centre
(\url{https://europeanspallationsource.se/data-management-software/computing-centre}). The
authors would like to thank Richard Heenan for his valuable comments on
the manuscript. Kalliopi Kanaki would like to thank Ramsey Al Jebali,
Morten Jagd Christensen and Jonas Nilsson for the useful conversations
and support with python.


\begin{thebibliography}{10}
\providecommand{\url}[1]{#1}
\csname url@samestyle\endcsname
\providecommand{\newblock}{\relax}
\providecommand{\bibinfo}[2]{#2}
\providecommand{\BIBentrySTDinterwordspacing}{\spaceskip=0pt\relax}
\providecommand{\BIBentryALTinterwordstretchfactor}{4}
\providecommand{\BIBentryALTinterwordspacing}{\spaceskip=\fontdimen2\font plus
\BIBentryALTinterwordstretchfactor\fontdimen3\font minus
  \fontdimen4\font\relax}
\providecommand{\BIBforeignlanguage}[2]{{%
\expandafter\ifx\csname l@#1\endcsname\relax
\typeout{** WARNING: IEEEtran.bst: No hyphenation pattern has been}%
\typeout{** loaded for the language `#1'. Using the pattern for}%
\typeout{** the default language instead.}%
\else
\language=\csname l@#1\endcsname
\fi
#2}}
\providecommand{\BIBdecl}{\relax}
\BIBdecl

\bibitem{esscdr}
S.~Peggs \emph{et~al.}, \emph{ESS Conceptual Design Report}, ESS 2012-001,
  \url{https://europeanspallationsource.se/sites/default/files/downloads/2017/09/CDR_final_120206.pdf}.

\bibitem{esstdr}
S.~Peggs \emph{et~al.}, \emph{ESS Technical Design Report}, ESS 2013-001,
  \url{https://europeanspallationsource.se/sites/default/files/downloads/2017/09/TDR_online_ver_all.pdf}.

\bibitem{kirstein2014}
O.~Kirstein \emph{et~al.}, \emph{Neutron Position Sensitive Detectors for the
  ESS, Proceedings of Science} \textbf{Vertex2014} (2014) 029.

\bibitem{lokijackson2015}
A.~J.~Jackson \emph{et~al.}, \emph{LoKI - A Broad Band High Flux SANS Instrument
  for the ESS, in Proceedings of ICANS XXI}, Mito, Ibaraki, Japan,
October 2015.

\bibitem{lokiproposal}
A.~J.~Jackson and K.~Kanaki, \emph{LoKI - A broad-band SANS instrument},
  \url{https://europeanspallationsource.se/sites/default/files/files/document/2017-09/loki_proposal_stc_sept2013.pdf},
  2012.

\bibitem{skadi2014}
S.~Jaksch \emph{et~al.}, \emph{Concept for a time-of-flight Small Angle Neutron Scattering instrument at the European Spallation Source,
  Nucl. Instr. Meth. A} \textbf{762} (2014) 22.

\bibitem{skadi2016}
S.~Jaksch \emph{et~al.}, \emph{Considerations about chopper configuration at a
  time-of-flight SANS instrument at a spallation source,
  Nucl. Instr. Meth. A} \textbf{835} (2016) 61.

\bibitem{sansbook}
L.~A.~Feigin and D.~Svergun, \emph{Structure Analysis by Small-Angle X-Ray and
  Neutron Scattering}, Springer (1987).

\bibitem{colin}
B.~T.~M.~Willis and C.~J.~Carlile, \emph{Experimental Neutron
  Scattering}, Oxford University Press (2009).

\bibitem{diffrrates}
I.~Stefanescu \emph{et~al.}, \emph{Neutron Detectors for the ESS
  diffractometers, J. Instr.}, \textbf{12} (2016) P01019.

\bibitem{mcstas1}
K.~Lefmann and K.~Nielsen, \emph{McStas, a general software package for neutron
  ray-tracing simulations}, \emph{Neutron News} \textbf{10} (1999) 20.

\bibitem{mcstas2}
P.~Willendrup, E.~Farhi, and K.~Lefmann, \emph{McStas 1.7 - a new version of the
  flexible Monte Carlo neutron scattering package, Physica B}
\textbf{350} (2004) E735.

\bibitem{essmoderator}
K.~Andersen \emph{et~al.}, \emph{Optimization of moderators and beam extraction at
  the ESS, J. Appl. Cryst.} \textbf{51} (2018) 04.

\bibitem{geant4a}
S.~Agostinelli \emph{et~al.}, \emph{Geant4: A Simulation toolkit, Nucl.
  Instrum. Meth. A} \textbf{506} (2003) 250.

\bibitem{geant4b}
J.~Allison \emph{et~al.}, \emph{Geant4 developments and applications,
  IEEE Trans. Nucl. Sci.} \textbf{53} (2006) 270.

\bibitem{geant4c_inpresscorrectedproof}
J.~Allison \emph{et~al.}, \emph{Recent developments in Geant4,
  Nucl. Instrum. Meth. A} \textbf{835} (2016) 07.

\bibitem{dgcode}
T.~Kittelmann \emph{et~al.}, \emph{Geant4 Based Simulations for Novel
  Neutron Detector Development, Journal of Physics: Conference
  Series}, \textbf{513} (2014) 022017. 

\bibitem{mcplpaper}
T.~Kittelmann \emph{et~al.}, \emph{Monte Carlo Particle Lists: MCPL,
  Computer Physics Communications} \textbf{218} (2017) 17.

\bibitem{mcplgithub}
\url{https://mctools.github.io/mcpl/},  \emph{MCPL documentation and GitHub
  repository} (accessed July 2018).

\bibitem{in6cncs}
E.~Dian \emph{et~al.}, \emph{Scattered neutron background in thermal neutron
  detectors, Nucl. Instr. Meth. A} \textbf{902} (2018) 173.

\bibitem{piscitelli2017}
F.~Piscitelli \emph{et~al.}, \emph{The Multi-Blade Boron-10-based neutron detector
  for high intensity neutron reflectometry at ESS, J. Instr.}
\textbf{12} (2017) P03013.

\bibitem{mgcncs}
A.~Khaplanov \emph{et~al.}, \emph{Multi-Grid Detector for Neutron Spectroscopy:
  Results Obtained on Time-of-Flight Spectrometer CNCS, J. Instr.}
\textbf{12} (2017) P04030.

\bibitem{lacy2013}
J.~L.~Lacy \emph{et~al.}, \emph{The evolution of neutron straw detector
  applications in homeland security, IEEE Trans. Nucl. Sci.}
\textbf{60} (2013) 1140.

\bibitem{sans1}
S.~M\"uhlbauer \emph{et~al.}, \emph{The new small-angle neutron scattering
  instrument SANS-1 at MLZ-characterization and first results, Nucl.
  Instr. Meth. A} \textbf{832} (2016) 297.

\bibitem{sans2d}
R.~K.~Heenan \emph{et~al.}, \emph{SANS2D at the ISIS Second Target
  Station, in Proceedings of ICANS-XVII} (2006) 780.

\bibitem{eqsans}
J.~K.~Zhao \emph{et~al.}, \emph{The extended Q-range small-angle neutron scattering
  diffractometer at the SNS, J. Appl. Cryst.} \textbf{43} (2010) 1068.

\bibitem{d33}
C.~D.~Dewhurst \emph{et~al.}, \emph{The small-angle neutron scattering instrument
  D33 at the Institut Laue-Langevin, J. Appl. Cryst.} \textbf{49} (2016) 1.

\bibitem{reuterstokes}
GE Digital Solutions,
\url{https://www.gemeasurement.com/radiation-measurement/neutron-scattering/helium-3-position-sensitive-neutron-detector}
(accessed July 2018).

\bibitem{toshiba}
Toshiba Electron Tubes and Devices,
\url{https://etd.canon/eng/product/prden.php?type=cat&search=500000300000}
(accessed July 2018).

\bibitem{illmultitube}
R.~A.~Campbell \emph{et~al.}, `\emph{FIGARO: The New Horizontal Neutron
  Reflectometer At The ILL, Eur. Phys. J. Plus} \textbf{126} (2011) 107.

\bibitem{nxsg4}
T.~Kittelmann and M.~Boin, \emph{Polycrystalline neutron scattering
  for Geant4: NXSG4, Computer Physics Communications}, \textbf{189} (2015) 114.

\bibitem{knollhe3}
G.~Knoll, \emph{Radiation Detection and Measurement, 4th Ed.}, John Wiley \&
  Sons, Inc., USA (2010) ch.~14, pg. 535--536, 539.

\bibitem{illbb}
ILL Neutron Data Booklet, \url{https://www.ill.eu/fileadmin/user_upload/ILL/1_About_ILL/Documentation/NeutronDataBooklet.pdf},
  2nd Edition, July 2003, ch. 3.3.8 (accessed July 2018).

\bibitem{boronstraw_lacy_2010}
J.~L.~Lacy \emph{et~al.}, \emph{Boron-coated straw detectors: A novel approach for
  helium-3 neutron detector replacement, in IEEE Nuclear Science
  Symposium and Medical Imaging Conference} (2010) 3971, doi:10.1109/NSSMIC.2010.5874561.

\bibitem{ncrystalgithub}
X.~X.~Cai and T.~Kittelmann, \emph{NCrystal : a library for thermal
  neutron transport in crystals},
\url{https://mctools.github.io/ncrystal/} (accessed July 2018)

\bibitem{icns}
K.~Kanaki \emph{et~al.}, \emph{Simulation tools for detector and
  instrument design, Physica B, in press} (2017) \url{https://doi.org/10.1016/j.physb.2018.03.025}.

\bibitem{nop_sonde}
S.~Jaksch \emph{et~al.}, \emph{Recent Developments SoNDe High-Flux Detector
  Project, JPS, in Proceedings of International Conference on Neutron
  Optics (NOP2017)} (2017).

\bibitem{sonde_patent}
Patent PCT/EP2015/074\,200, \emph{Scintillation detector with a high
  count rate} (2015).

\bibitem{sonde_arxiv}
S.~Jaksch \emph{et~al.}, \emph{Cumulative Reports of the SoNDe
  Project, arXiv:1707.08679} (2017).

\bibitem{sonde_web}
The SoNDe project at FZJ,
\url{http://www.fz-juelich.de/jcns/jcns-2/EN/Forschung/Instruments-for-ESS/SoNDe-Projekt/_node.html}
(accessed July 2018).

\bibitem{gs20}
GS20 $^6$Li-glass specifications, \url{https://scintacor.com/products/6-lithium-glass} (accessed July 2018).

\bibitem{12700}
MaPMT H12700 specifications by Hamamatsu Photonics,
\url{https://www.hamamatsu.com/resources/pdf/etd/H12700_TPMH1348E.pdf}
(accessed July 2018).

\bibitem{bandgem1}
G.~Croci \emph{et~al.}, \emph{Diffraction measurements with a boron-based GEM
  neutron detector, EPL (Europhysics Letters)} \textbf{107} (2014) 12001.

\bibitem{bandgem2}
E.~Perelli~Cippo \emph{et~al.}, \emph{A GEM-based thermal neutron detector for high
  counting rate applications, J. Instr.} \textbf{10} (2015) P10003.

\bibitem{bandgem3}
G.~Albani \emph{et~al.}, \emph{Evolution in boron-based GEM detectors for
  diffraction measurements: from planar to 3D converters, Measurement
  Science and Technology} \textbf{27} (2016) 115902.

\bibitem{bandgem4}
A.~Muraro \emph{et~al.}, \emph{Performance of the high-efficiency thermal neutron
  BAND-GEM detector, Prog. Theor. Exp. Phys.} \textbf{023H01} (2018).

\bibitem{giorgiaphd}
G.~Albani, \emph{High rate thermal neutron gaseous detector for use at neutron
  spallation sources, Ph.D. dissertation}, Dipartimento di Fisica ``G.
  Occhialini'', Universit\`a degli Studi di Milano-Bicocca (2017).

\bibitem{9500}
MaPMT H9500 specifications by Hamamatsu Photonics,
\url{https://www.hamamatsu.com/resources/pdf/etd/H9500_H9500-03_TPMH1309E.pdf}
(accessed July 2018).

\bibitem{b10nano}
F.~D.~Amaro \emph{et~al.}, \emph{Novel concept for neutron detection:
  proportional counter filled with $^{10}$B nanoparticle aerosol,
  Scientific Reports 7:41699} (2017).

\bibitem{thickGEM}
M.~Cortesi \emph{et~al.}, \emph{Development of a cold-neutron imaging
  detector based on thich gaseous electron multiplier,
  Rev. Sci. Instrum.} \textbf{84} (2013) 023305.

\bibitem{b10suspension}
K.~A.~Nelson \emph{et~al.}, \emph{A suspended boron foil multi-wire
  proportional counter neutron detector, Nucl. Instrum. Meth. A},
\textbf{767} (2014) 14.

\end{thebibliography}


\end{document}